\documentclass[twocolumn,nofootinbib,amsmath,amssymb]{revtex4}
\usepackage{graphicx}

\newcommand{\e}{\mathrm{e}}
\newcommand{\Imag}{\mathop{\mathrm{Im}}}
\newcommand{\Real}{\mathop{\mathrm{Re}}}
\newcommand{\bra}[1]{\langle #1 |}
\newcommand{\ket}[1]{| #1 \rangle}

\begin{document}

\preprint{INR/TH-2003-1}

\title{Dynamical tunneling of bound systems through a potential barrier: 
complex way to the top.}

\author{F.~Bezrukov}
\email{fedor@ms2.inr.ac.ru}
\affiliation{
  Institute for Nuclear Research of the Russian Academy of Sciences,\\
  60th October Anniversary prospect 7a, Moscow 117312, Russia}
\author{D.~Levkov}
\email{levkov@ms2.inr.ac.ru}
\affiliation{
  Institute for Nuclear Research of the Russian Academy of Sciences,\\
  60th October Anniversary prospect 7a, Moscow 117312, Russia}
\affiliation{
  Moscow State University, Department of Physics,\\
  Vorobjevy Gory, Moscow, 119899, Russian Federation}
\date{January 7, 2003}

\begin{abstract}
A semiclassical method of complex trajectories for the calculation of the
tunneling exponent in systems with many
degrees of freedom is further developed. It is supplemented with an
easily implementable technique, which enables one to single out the 
physically relevant trajectory from the whole set of complex classical 
trajectories. The method is applied to
semiclassical transitions of a bound system through a potential
barrier. We find that the properties of physically relevant 
complex trajectories are
qualitatively different in the cases of potential tunneling at low energy
and dynamical tunneling at energies exceeding the barrier height. 
Namely, in the case of high
energies, the physically relevant complex trajectories describe tunneling
via creation of a state close to the top of the barrier. The method is 
checked against exact solutions of the Schr\"odinger equation in a quantum 
mechanical system of two degrees of freedom.
\end{abstract}


\newpage
\maketitle

\onecolumngrid
\tableofcontents
\twocolumngrid
\mbox{}
\cleardoublepage

\section{\label{sec:intro}Introduction}

Semiclassical methods provide a useful tool for the study of
nonperturbative processes.  Tunneling phenomena represent one of the
most notable cases where semiclassical techniques are used to
obtain otherwise unattainable information on the dynamics of the transition. 
A standard example of semiclassical technique is WKB approximation
to tunneling in quantum mechanics of one 
degree of freedom.
In this case solutions $S(q)$ to the Hamilton--Jacobi equation are
pure imaginary in the classically forbidden region. 
Therefore, the function $S(q)$ can be obtained as the action functional on 
a {\it real} trajectory $q(\tau)$, which is the solution to the 
equations of motion in 
Euclidean time domain, $t = -i\tau$, with real Euclidean action
$S_E = -i S$. 

This simple picture of tunneling is no longer valid for systems with many 
degrees of freedom, where solutions $S(\mathbf{q})$ to the Hamilton--Jacobi 
equation are known to be generically complex in the classically forbidden 
region (see Refs.~\cite{Huang:1989,Takada:1993} for recent discussion). 
This leads to the concept of ``mixed'' tunneling, as opposed to ``pure'' 
tunneling where $S(\mathbf{q})$ is pure imaginary. 
``Mixed'' tunneling cannot be described by any real tunneling trajectory.
However, it could be related to a {\it complex} trajectory. If so,
the function $S(\mathbf{q})$ (and therefore the exponential part of the 
wave function) is calculated as the action functional on this complex 
trajectory. 

A particularly difficult situation arises when one considers transitions of a 
non--separable system with a strong interaction between its degrees 
of freedom, such that the quantum numbers of the system change considerably 
during the transition. Methods based on adiabatic expansion are not applicable
in this situation, while the method of complex trajectories proves to be 
extremely useful.

The method of complex trajectories in the form suitable for the calculation
of $S$--matrix elements was formulated and checked by direct numerical 
calculations in Refs.~\cite{Miller:1970,Miller:1972,George:1972} 
(see Ref.~\cite{Miller} for review). Further 
studies~\cite{Wilkinson:1986,Wilkinson:1987,Takada:1995,Takada:1996,
Bonini:1999cn,Bonini:1999kj} showed
that this method can be generalized to the  calculation of the tunneling wave 
functions and tunneling probabilities, energy splittings in double well 
potentials and decay 
rates from metastable states. Similar methods were successful in the
study of tunneling in high-energy collisions in field theory
\cite{Rubakov:1992ec,Kuznetsov:1997az,Bezrukov:2001dg,Bezrukov:2003er}, 
where one  
considers systems with definite particle number (${\cal N} = 2$) in the 
initial state; in the 
study of chemical reactions and atom ionization processes, where the initial 
bound systems are in definite quantum 
states~\cite{Miller,Perelomov1,Perelomov2}; etc. 
The main advantage of the method of complex trajectories
is that it can be easily generalized and numerically implemented in the cases
of large and even 
infinite (field theory) number of degrees of freedom, in 
contrast  to other methods such as Huygens-type 
construction of Refs.~\cite{Huang:1989,Takada:1993} and  initial value 
representation (IVR) of Refs.~\cite{Miller:1970,Miller:1988,Thompson:1999,
Kay:1997,Maitra:1997,Miller:2001}.

In this paper we develop the method of complex trajectories further.
Namely, we concentrate on the following problem. It is
known~\cite{Miller:1970} that the physically relevant complex trajectory 
satisfies the classical equations of motion with certain boundary 
conditions. However, this boundary value problem  generically has also 
an infinite, 
though discrete, set of unphysical solutions.
In one dimensional quantum mechanics all solutions
can easily be classified. In systems with many 
degrees of freedom such a classification  is extremely difficult, if at
all possible.
In the case of small number of degrees of freedom (realistically, $N = 2$),
one can scan over all solutions and find the solution giving the
largest tunneling probability~\cite{Miller:1970,Takada:1995,Takada:1996}, but in systems with large or infinite number of 
degrees of freedom the problem of choosing the physically relevant
solution becomes a formidable task.

The problem of choosing the appropriate solution becomes even more 
pronounced when the qualitative properties of the relevant complex trajectory
are different in different energy regions.
This may happen when the physically relevant classical solution ``meets''
an unphysical one at some value of energy $E = E_1$, or in other words,
when solutions to the boundary
value problem, viewed as functions of energy, bifurcate at $E = E_1$. 

In this paper we give an example of this sort, which appears to be fairly 
generic (see also~\cite{Bonini:1999cn,Bonini:1999kj,Kuznetsov:1996cm,
Bezrukov:2001dg,Bezrukov:2003er}). We then develop a method which 
chooses the phyically relevant solution automatically,
implement it numerically
and check this method against the numerical
solution to the full Schr\"odinger equation. 

We study inelastic transitions 
of a bound system through a potential barrier. 
For concreteness we consider a model with one internal degree of freedom
besides the center-of-mass coordinate.
We consider a situation in which 
the spacing between the levels of the  bound system is small compared 
to the height of the barrier, and assume strong enough coupling between the
degrees of freedom, to make sure that the quantum numbers of the
bound system change considerably during the transition process.
This is precisely the situation in which the method of complex trajectories
shows its full strength.
 
Transitions of bound systems involve a particular energy 
scale --- the height of the barrier $V_0$. At energies below $V_0$ classical 
over--barrier transitions are forbidden energetically; the corresponding 
regime is called ``potential tunneling''. 
For $E >  V_0$ it is energetically allowed for the system to evolve 
classically to the other side of the barrier. However, over-barrier 
transitions  may be forbidden  dynamically even at $E > V_0$. Indeed, 
inelastic interactions of a bound system with a
potential barrier generally lead 
to the excitation of the internal degrees of freedom with the simultaneous
decrease of the center-of-mass energy, and this may  prevent the system from 
the over--barrier transition. Tunneling regime at energies exceeding the 
barrier height is called ``dynamical tunneling''\footnote{
It is clear that the properties of transitions of a bound system at 
$E > V_0$ depend on the choice of the initial state. Namely, there 
always exists a certain class of states, transitions from which are not 
exponentially suppressed.To construct an example, one places the bound system 
on top of the barrier and evolves it classically backwards in time to the 
region where the interaction with the barrier is negligibly small.
On the other hand, even at $E > V_0$ there are 
states, transitions from which are exponentially suppressed (dynamical 
tunneling). 
}.

Examples of dynamical tunneling are well--known in scattering 
theory~\cite{Miller:1972}. This type of tunneling between bound states
was discovered in Ref.~\cite{Heller:1981}, 
the generality of dynamical tunneling 
in large molecules was stressed in Refs.~\cite{Davis:1981,Heller:1994}. 
It is dynamical tunneling that is of primary interest in our study.

A novel phenomenon we observe is that dynamical 
tunneling at $E \gtrsim V_0$ (more precisely, at 
$E > E_1$, where $E_1$ is somewhat larger than $V_0$) occurs in the following
way: the system jumps on top of the barrier, and restarts its classical 
evolution from the region near the top. 
 From the physical viewpoint, this is not quite what is
normally meant by ``tunneling through a barrier''. Yet the transitions
remain exponentially suppressed, but the reason is different: to
jump above the barrier, the system has to undergo considerable
rearrangement, unless the incoming state is chosen in a special way
(see footnote, above).
This rearrangement costs exponentially small probability factor. We
note that similar exponential factor was argued to appear in various
field theory processes with multi-particle final
states~\cite{Banks:1990zb,Zakharov:1991rp,Veneziano:1992rp,Rubakov:1995hq}.

We find that the new physical behaviour of the system is related to a 
bifurcation
of the family of the complex-time classical solutions, viewed as functions of 
energy. This is precisely the bifurcation which we alluded to above.
Our method of dealing with this bifurcation is to regularize the boundary value
problem in such a way that the bifurcations disappear altogether (at real
energies), and the only solutions recovered after removing the regularization
are physical ones.

The paper is organized as follows. The system we discuss in this paper
is introduced in Sec.~\ref{sec:2Dsystem}.
In Sec.~\ref{sec:Ttheta} we formulate
the boundary value problem for the calculation of the
tunneling exponent. Then we examine the classical over-barrier
solutions and find all initial states that lead to classically allowed
transitions in Sec.~\ref{sec:over-barrier}.  In
Sec.~\ref{sec:bifurcation} we present a straightforward application of
the semiclassical technique, outlined in Sec.~\ref{sec:Ttheta}, and
find that it ceases to produce relevant complex trajectories 
n a certain region of initial data, namely, at $E>E_1$.
In Sec.~\ref{sec:4} we introduce our regularization technique and show
that it indeed enables one to find all the relevant complex trajectories,
including ones with $E > E_1$ (Sec.~\ref{sec:reg_forb}).  We check our method
against the numerical solution of the full Schr\"odinger equation in
Sec.~\ref{sec:4.2}.  In Sec.~\ref{sec:4.3} and Appendix~\ref{sec:1D} 
we show how our
regularization technique is  used to join smoothly the ``classically
allowed'' and ``classically forbidden'' families of solutions in the cases 
of two-  and one- dimensional quantum mechanics, respectively.

\section{\label{sec:2D}Semiclassical transitions through a potential barrier}
\subsection{\label{sec:2Dsystem} The model}

\begin{figure}
  \centerline{\includegraphics[width=0.8\columnwidth]{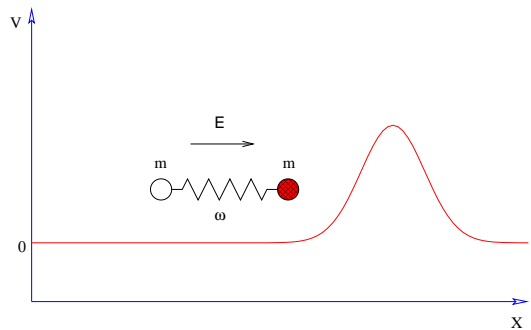}}
  \caption{Oscillator hitting a potential barrier, with which only the
 ``dark'' particle interacts.}
  \label{fig0}
\end{figure}

The situation we discuss in this paper is a transition 
through a potential barrier of a bound system of 
Refs.~\cite{Bonini:1999cn,Bonini:1999kj}, namely a system made of
two particles of identical mass $m$, moving in one dimension and bound
by a harmonic oscillator potential of frequency $\omega$ (Fig.~\ref{fig0}).
One of the
particles interacts with a repulsive potential barrier. The potential barrier
 is
assumed to be high and wide, while spacing between the oscillator
levels is much smaller than the barrier height $V_0$.   
The Hamiltonian of the model is
\begin{equation}\label{Hold}
  H = \frac{p_1^2}{2m}+\frac{p_2^2}{2m}+\frac{m\omega^2}{4}(x_1-x_2)^2
      +V_0\e^{- x_1^2/2\sigma^2} \;,
\end{equation}
where the conditions on the oscillator frequency and potential barrier are
\begin{eqnarray}
  &&\hbar\omega\ll V_0, \label{conditions}\\
  &&\sigma\gg\hbar/\sqrt{mV_0}\;.\nonumber
\end{eqnarray}
Since the variables do not separate, this is certainly a non-trivial system.

We choose  units with 
$\hbar = 1,$ $m = 1$. It is also convenient to treat the frequency $\omega$ as
 a dimensionless parameter, so that all physical quantities 
are dimensionless. In our subsequent numerical study we use the value
$\omega=0.5$, still keeping, however, notation ``$\omega$'' in formulas.
The system is  semiclassical, i.e.\  conditions
\eqref{conditions} are satisfied, if one chooses 
$\sigma = 1/\sqrt{2 \lambda}$, $V_0 =1/\lambda$, where $\lambda$ 
is a small parameter. At the classical level, this parameter is irrelevant:
after rescaling the variables\footnote{To keep notations simple, we use the 
same symbols $x_1,\;x_2$ for the rescaled variables.}
$x_1 \to x_1/\sqrt{\lambda},\;\; x_2 \to  x_2/\sqrt{\lambda},$
the small parameter enters only through the overall multiplicative factor
$1/\lambda$ in the Hamiltonian. 
Therefore, the semiclassical technique can
be developed as an asymptotic expansion in $\lambda$. 

The properties of the system are
made clearer by replacing the variables $x_1,\;x_2$ with the
center-of-mass coordinate $ X \equiv (x_1+x_2)/\sqrt{2}$ and the
relative oscillator coordinate $y\equiv (x_1 - x_2)/\sqrt2$. 
In terms of the latter variables, the Hamiltonian takes the form
\begin{equation}
  H = \frac{p_X^2}{2} + \frac{p_y^2}{2} + \frac{\omega^2}{2} y^2 +
  \frac1\lambda \mathrm{e}^{- \lambda (X+y)^2/2}.
  \label{H}
\end{equation}
The interaction potential
\begin{equation*}
  U_{\mathrm{int}} \equiv \frac{1}{\lambda} \mathrm{e}^{- \lambda (X+y)^2/2}
\end{equation*}
vanishes in the asymptotic regions $X \to \pm\infty$ and describes a 
potential barrier between these regions. At $X\to \pm\infty$ the 
Hamiltonian~\eqref{H} corresponds to an 
oscillator of frequency $\omega$ moving along the center-of-mass coordinate 
$X$. The oscillator asymptotic state is characterized by 
its excitation number $N$  and total energy 
$ E = p_X^2/2 + \omega(N + 1/2)$. 
We are interested in the transmissions through the potential barrier of the 
oscillator with given initial values of $E$ and $N$.

\subsection{\label{sec:Ttheta}$T/\theta$ boundary value problem}

The probability of tunneling from a
state with fixed initial energy $E$ and oscillator excitation number $N$
from the asymptotic region $X \to -\infty$ to any state in the other
asymptotic region $X\to+\infty$ takes the following form:
\begin{equation}\label{TT}
  \mathcal{T}(E,N) = \lim_{t_f - t_i \to \infty}\sum_{f} 
  \left|\bra{f} \e^{- i \hat{H}(t_f-t_i)} \ket{E,N}\right|^2
  \;,
\end{equation}
where it is implicit
that the initial and final states have support only well outside 
the range of the potential, with $X<0$ and $X>0$, respectively.
Semiclassical methods are applicable when the initial energy and excitation 
number are parametrically large,
$$
  E = \tilde{E}/\lambda
  \;,\quad
  N = \tilde{N}/\lambda
  \;,
$$
where $\tilde{E}$ and $\tilde{N}$ are held constant as $\lambda\to 0$.
The transition probability has the exponential form
\begin{equation}
\label{+}
{\cal T} = D \mathrm{e}^{-\frac{1}{\lambda} F(\tilde{E},\;\tilde{N})}\;,
\end{equation}
where $D$ is a pre-exponential factor, which is not considered in
this paper.  Our purpose is to calculate the leading semiclassical
exponent $F(\tilde{E},\; \tilde{N})$. The exponent for tunneling from
the oscillator ground state is 
obtained~\cite{Rubakov:1992fb,Rubakov:1992ec,Bonini:1999cn,Bonini:1999kj} 
by taking the limit ${\tilde{N}\to 0}$ in $F(\tilde{E},\tilde{N})$.

\begin{figure}
  \centerline{\includegraphics[width=\columnwidth]{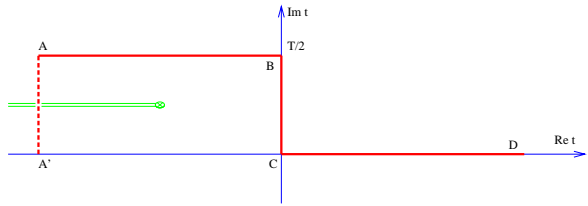}}
  \caption{Contour in the complex time plane.}
\label{fig2}
\end{figure}
In what follows we rescale the variables, $X\to X/\sqrt{\lambda}$, 
$y\to y/\sqrt{\lambda}$, and omit tilde over the
rescaled quantities $\tilde{E},\;\tilde{N}$.

The exponent $F(E,N)$ is related to a complex trajectory, which satisfies
a certain complexified classical boundary value problem. We present 
the derivation of this problem 
in Appendix~\ref{app:A}.  The outcome is as follows. There are two Lagrange 
multipliers, $T$ and $\theta$,
which are related to the parameters $E$ and $N$ characterizing the
incoming state.   The boundary
value problem is conveniently formulated on the contour ABCD in the
complex time plane (see Fig.~\ref{fig2}), with the imaginary part of
the initial time equal to $T/2$.  The coordinates $X(t),\; y(t)$ must
satisfy the complexified equations of motion in the internal points of
the contour, and be real in the asymptotic future (region D):
\begin{subequations}\label{bc}
\begin{eqnarray}
  && \frac{\delta S}{\delta X(t)}=\frac{\delta S}{\delta y(t)} = 0
  \;,\\[1ex]
  &&
  \begin{array}{l}
    \displaystyle \Imag y(t) \to 0, \\[1ex]
    \displaystyle \Imag X(t) \to 0,  
  \end{array}
  \qquad\text{as } t \to +\infty \;.
  \label{bcplus}
\end{eqnarray}
In the asymptotic past (region A of the contour, where $t = t' + i
T/2$, $t'$ is real negative) one can neglect the interaction potential
$U_\mathrm{int}$, and the oscillator decouples:
\begin{equation*}
  y = \frac{1}{\sqrt{2 \omega}}(u \mathrm{e}^{- i \omega t'} +  
  v \mathrm{e}^{i \omega t'}).
\end{equation*}
The boundary conditions in the asymptotic past, $t' \to -\infty$, are
that the center-of-mass coordinate $X$ must be real, while the complex
amplitudes of the decoupled oscillator must be linearly related,
\begin{equation}\label{bcminus}
\begin{array}{l}
\Imag X \to 0,\\[1ex]
  v \to \e^\theta u^*,\; 
\end{array}
  \qquad  \text{as } t' \to -\infty\;.
\end{equation}
\end{subequations}
The boundary conditions \eqref{bcplus} and \eqref{bcminus} make, in
fact, eight real conditions (since, e.g.,\ $\Imag X(t')\to0$ implies
that both $\Imag X$ and $\Imag\dot{X}$ tend to zero), and completely
determine a solution, up to time translation invariance (see
discussion in Appendix~\ref{app:A}).

It is shown in Appendix~\ref{app:A} that a solution to this boundary value
problem is an extremum of the functional
\begin{multline}\label{F}
  F[X,y;X^*,y^*;T,\theta] = - iS[X,y] + i S[X^*,y^*] \\
    - ET - N\theta
    + \text{Boundary terms}\;.
\end{multline}
The value of this functional at the extremum gives the exponent for the
 transition probability (up to large overall factor $1/\lambda$, 
see eq.~\eqref{+}),
\begin{equation}\label{Fshort}
  F(E,\;N) = 2 \Imag  S_0(T,\;\theta) - E T - N \theta\;,
\end{equation}
where $S_0$ is the action of the solution, integrated by parts,
\begin{multline}\label{8*}
  S_0 = \int \! dt\left(\!
    - \frac12 X\frac{d^2 X}{dt^2} - \frac12 y \frac{d^2 y}{dt^2} \right.\\ 
    \left. - \frac12 \omega^2 y^2 - U_{\mathrm{int}}(X,y)
  \right).
\end{multline}
Here the integration runs along the contour ABCD.
The values of the Lagrange multipliers $T$ and $\theta$ are related 
to energy and excitation number as follows,
\begin{eqnarray}
  E(T,\theta) &=& \frac\partial{\partial T} 2 \Imag  S_0(T,\theta)\;,
  \label{E}\\
  N(T,\theta) &=& \frac\partial{\partial \theta} 2 \Imag  S_0(T,\theta)\;.
  \label{N}
\end{eqnarray}
Making use of Eq.~\eqref{Fshort}, it is straightforward to check also the
inverse Legendre transformation formulas,
\begin{eqnarray}
  T(E,\;N) &=& - \frac{\partial}{\partial E} F(E,\;N) \;,
  \label{T}\\
  \theta(E,\;N) &=& - \frac{\partial}{\partial N} F(E,\;N) \;.
  \label{theta}
\end{eqnarray} 
One can also check that the right hand side of Eq.~\eqref{E} coincides with
the energy of the classical solution, while the right hand side of
Eq.~\eqref{N} is equal to the classical counterpart of the occupation
number,
\begin{equation}\label{energyeq}
  E=\frac{\dot{X}^2}{2}+\omega N
  \;;\qquad
  N = uv\;.
\end{equation}
So, one may either search for the  values of $T$ and
$\theta$ that correspond to given $E$ and $N$, or, following a
 computationally simpler procedure, solve the boundary
value problem \eqref{bc} for given $T$ and $\theta$ and then find the
corresponding values of $E$ and $N$ from
Eq.~\eqref{energyeq}. Note that the initial conditions~\eqref{bcminus}
complemented by Eqs.~\eqref{energyeq} are equivalent to the initial
conditions of Refs.~\cite{Miller:1970,Miller:1972,George:1972}, the latter 
being expressed in terms of action--angle variables.
The boundary conditions in the asymptotic future~\eqref{bcplus} are different
from those of Refs.\cite{Miller:1970,Miller:1972,George:1972}, since we
consider inclusive, rather  than fixed, final state.

Let us discuss some subtle points of the boundary value problem
\eqref{bc}. First, one notices that the condition of asymptotic
reality \eqref{bcplus} does not always coincide with the condition of
reality at finite time.  Of course, if the solution approaches the
asymptotic region $X\to+\infty$ on the part CD of the contour, the
asymptotic reality condition \eqref{bcplus} implies that the solution
is real at any \emph{finite} positive $t$.  Indeed, the oscillator
decouples as $X\to+\infty$, so the condition \eqref{bcplus} means that
its phase and amplitude, as well as $X(t)$, are real as $t\to
+\infty$.  Due to equations of motion, $X(t)$ and $y(t)$ are real on
the entire CD--part of the contour.  This situation corresponds to the
transition directly to the asymptotic region $X\to+\infty$.  However,
the situation can be drastically different if the solution on the
final part of the time contour remains in the interaction region.  For
example, let us imagine that the solution approaches the saddle point
of the potential $X = 0$, $y = 0$ as $t\to+\infty$.  Since one of the
perturbations about this point is unstable, there may exist solutions
which approach this point {\it exponentially} along the unstable
direction, i.e.\ $X(t),\; y(t)\propto \mathrm{e}^{- \mathrm{const}
\cdot t}$ with possibly complex pre-factors.  In this case the
solution may be complex at any finite time, and become real only
asymptotically, as $t\to+\infty$.  Such solution corresponds to
tunneling to the saddle point of the barrier, after which the system
rolls down classically towards $X\to+\infty$ (with probability of
order $1$, inessential for the tunneling exponent $F$).  We will see
in Sec.~\ref{sec:reg_forb} that the situation of this sort indeed
takes place for some values of energy and excitation number.

Second, since at large negative time (in the asymptotic region
$X\to-\infty$) the interaction potential disappears, it is
straightforward to continue the asymptotics of the solution to the
real time axis.  For solutions satisfying \eqref{bcminus} this gives
at large negative time
\begin{eqnarray}
&&  y(t) = \frac{1}{\sqrt{2\omega}}\left(
    u \e^{-\omega T/2}\e^{-i\omega t} +  
    u^*\e^{\theta+\omega T/2}\e^{i\omega t}
  \right) \;,\nonumber
  \\
&&  \Imag X(t) = -\frac{T}{2}p_X \;.\nonumber
\end{eqnarray}
We see that the dynamical coordinates on the negative side of the real
time axis are generally complex.  For solutions approaching the
asymptotic region $X\to+\infty$ as $t\to +\infty$ (so that $X$ and $y$
are exactly real at finite $t>0$), this means,
that there should exist a branch point in
the complex time plane: the contour A'ABC in Fig.~\ref{fig2} winds around
this point and cannot be deformed to the real time axis.  This
argument \emph{does not} work for solutions ending in the interaction
region as $t\to+\infty$, so branch points between the AB--part of the
contour and the real time axis may be absent.  We will see in
Sec.~\ref{sec:reg_forb} that this is indeed the case in our model in
a certain range of $E$ and $N$.

\subsection{\label{sec:over-barrier}Over-barrier transitions: the region of
classically allowed transitions and its boundary $E_0(N)$}

Before studying the exponentially suppressed transitions, let us
consider the classically allowed ones. To this end, let us study the
classical evolution (real time, real-valued coordinates), in which the system is initially located at large
negative $X$ and moves with positive center-of-mass velocity towards
the asymptotic region $X \to +\infty$.  The classical dynamics of the
system is specified by four initial parameters.  One of them (e.g.,
the initial center-of-mass coordinate) fixes the invariance under time
translation, while the other three are the total energy $E$, the
initial excitation number of the $y$--oscillator, defined in classical
theory as $N\equiv E_\mathrm{osc}/\omega$, and the initial oscillator
phase $\varphi_i$. 

Any initial quantum state of our system can be fully determined by energy
$E$ and initial oscillator excitation number $N$; we can represent each state
by a point in the $E$--$N$ plane. There is, however, one additional
classically relevant initial parameter, the oscillator phase~$\varphi_i$.
An initial state $(E,N)$ leads to unsuppressed transmission
if the corresponding classical over--barrier transitions\footnote{Note that the corresponding classical solutions 
obey the boundary conditions~\eqref{bcplus}, \eqref{bcminus} with 
$T = \theta = 0$, i.e., they are solutions to the boundary value 
problem~\eqref{bc}.} are possible for {\it some} value(s) of
$\varphi_i$. These states form some region 
in the $E$--$N$ plane, which is to be found in this section. 

\begin{figure}
  \centerline{\includegraphics[width=\columnwidth]{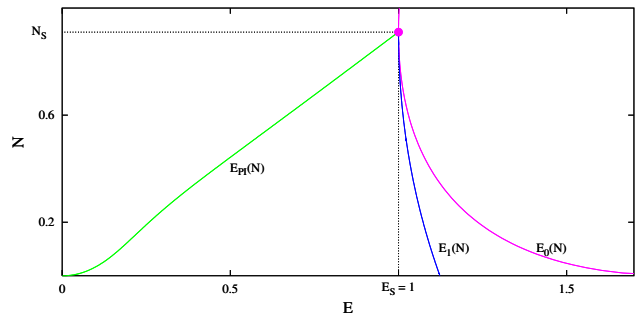}}
  \caption{The boundary~$E_0(N)$ of 
    the region of classically allowed transitions, 
    the bifurcation line $E_1(N)$ and the
    line of the periodic instantons $E_{PI}(N)$.}
  \label{fig7} 
\end{figure}

For given $N$, at large enough $E$ the system can certainly evolve to the 
other side of the barrier.  On the other
hand, if $E$ is smaller than the barrier height, the system
definitely undergoes reflection.  Thus, there exists some boundary
energy $E_0(N)$, such that classical
transitions are possible for $E > E_0(N)$, while for
$E < E_0(N)$ they do not occur for \emph{any} initial phase
$\varphi_i$.  The line $E_0(N)$ represents the boundary of the region of
classically allowed transitions.
We have calculated $E_0(N)$ numerically: the result\footnote{Note that 
the boundary $E_0(N)$ of the region of classically allowed transitions 
can be extended to $N > N_{\mathrm{S}}$. As 
$E = E_{\mathrm{S}}$ is the absolute minimum of the energy of classically 
alowed transitions, the function $E_0(N)$ grows 
with $N$ at $N > N_{\mathrm{S}}$. In fact,
it tends to asymptotics $E_0^{\mathrm{as}} = \omega N$ as $N\to +\infty$.
In what follows we are not interested in transitions with $N > N_{\mathrm{S}}$,
so this part of the boundary $E_0(N)$ is not presented in 
Fig.~\ref{fig7}} is 
shown in Fig.~\ref{fig7}.

\begin{figure}
  \begin{center}
    \includegraphics[width=1.1\columnwidth]{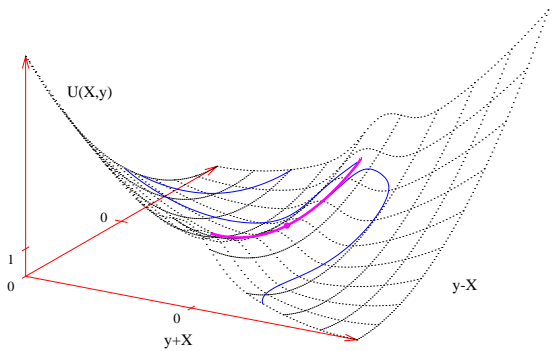}
  \end{center}
  \caption{The potential (dotted lines)
    in the vicinity of the sphaleron ${(X = 0,\;y = 0)}$ (marked by
    the point), excited sphaleron (thick line)
    corresponding to the point $(E,N) = (1.985,3.72)$.
    at the boundary of the region of classically allowed transitions, and the
    trajectory of the solution which is close to this excited
    sphaleron (thin line).  In this figure the asymptotic regions
    $X\to\pm\infty$ are along the diagonal.}
  \label{fig8} 
\end{figure}

An important point of the boundary $E_0(N)$ corresponds to the static
unstable classical solution $X(t) = y(t) = 0$. In field theory context, 
such a solution is called 
``sphaleron''~\cite{Klinkhamer:1984di}),
and for the sake of terminology we will use this name in what follows.
This solution is the saddle point
of the potential $U(X,y)\equiv {\omega^2
y^2}/{2}+U_{\mathrm{int}}(X,y)$ and has exactly one unstable
direction, the negative mode (see Fig.~\ref{fig8}).  The sphaleron
energy $E_\mathrm{S} = U(0,0) = 1$ determines the minimum value of the
function $E_0(N)$.  Indeed, while classical over--barrier transitions
with ${E < E_{\mathrm{S}}}$ are impossible, the over--barrier solution
with slightly higher energy can be obtained as follows: at the point
$X=y=0$, one adds momentum along the negative mode, thus ``pushing''
the system towards $X\to+\infty$.  Continuing this solution backwards
in time one finds that the system tends to $X\to-\infty$ for large
negative time and has a certain oscillator excitation number.
Solutions with energy closer to the sphaleron energy correspond to
smaller ``push'' and thus spend longer time near the sphaleron.  In
the limiting case when the energy is equal to $E_{\mathrm{S}}$, the
solution spends an infinite time in the vicinity of the sphaleron.
This limiting case has a definite initial excitation number
$N_{\mathrm{S}}$, so that $E_0(N_{\mathrm{S}}) = E_\mathrm{S}$ (see
Fig.~\ref{fig7}).  The value of $N_\mathrm{S}$ is unique because there
is exactly one negative direction of the potential in the vicinity of
the sphaleron.

In complete analogy to the features of the over--barrier classical
solutions near the sphaleron point ($E_\mathrm{S}$, $N_{\mathrm{S}}$),
one expects that as the values of $E,\;N$ approach any other boundary
point $(E_0(N),\;N)$, the corresponding over--barrier solutions will
spend more and more time in the interaction region, where
$U_{\mathrm{int}} \ne 0$.  This follows from a continuity
argument.  Namely, let us first fix the initial and final times, $t_i$
and $t_f$.  If in this time interval a solution with energy $E_1$
evolves to the other side of the barrier and a solution with energy
$E_2$ and the same oscillator excitation number is reflected back,
there exists an intermediate energy at which the solution ends up at
$t = t_f$ in the interaction region.  Taking the limit $t_f \to
+\infty$ and $(E_1-E_2)\to0$, we obtain a point at the boundary $E_0(N)$ 
and a solution tending asymptotically to
some unstable time dependent solution that spends infinite time in the
interaction region.  We call the latter solution \emph{excited
sphaleron;} it describes some (in general nonlinear) oscillations
above the sphaleron along the stable direction in coordinate space.
Therefore, every point of the boundary ${(E_0(N),\;N)}$ corresponds to
some excited sphaleron. Solutions tending asymptotically to the excited 
sphalerons, form a surface in a phase space (separatrix), 
which separates regions of qualitatively different classical motion of the 
system. 

We display in Fig.~\ref{fig8} a solution, found numerically in our
model, that tends to an excited sphaleron.  We see that the
trajectory of excited sphaleron is, roughly speaking, 
orthogonal to the unstable direction at the saddle point $({X = 0,}\;{y = 0})$.

\subsection{\label{sec:bifurcation}Suppressed transitions: bifurcation
    line $E_1(N)$}

Let us now turn to classically forbidden transitions, and consider the 
boundary value problem~\eqref{bc}. It is relatively straightforward to 
obtain numerically  solutions for $\theta = 0$.
The boundary conditions
\eqref{bcplus}, \eqref{bcminus} in this case take the form of 
reality conditions in the asymptotic future and past. It can be 
shown~\cite{Khlebnikov:1991th} that the physically relevant solutions
with $\theta = 0$ are real on the entire contour ABCD of Fig.~\ref{fig2} and 
describe nonlinear oscillations in the
upside-down potential on the Euclidean part BC of the contour. The period of 
the
oscillations is equal to $T$, so that the points B and C are two different 
turning
points, where $\dot{X} = \dot{y} = 0$.  These real Euclidean
solutions are called periodic instantons. A practical technique for obtaining these 
solutions numerically on the Euclidean part BC consists in minimizing
 the Euclidean action (for example, with the method of 
conjugate gradients, see Ref.~\cite{Bonini:1999cn,Bonini:1999kj} for details). 
The solutions on the entire contour are then obtained by solving numerically 
the Cauchy problem, forward in time along the line CD
and backward in time along the line BA. Having the solution in asymptotic past
(region A), one then calculates its energy and excitation
number~\eqref{energyeq}. The solutions to this Cauchy problem are obviously real,
so the boundary conditions~\eqref{bcplus}, \eqref{bcminus} are indeed 
satisfied with $\theta = 0$. It is worth noting that solutions with $\theta = 0$
are similar to the ones in quantum mechanics of one degree of freedom.
The line of 
periodic instantons in $E$--$N$ plane in our model is
shown in Fig.~\ref{fig7}.

Once the solutions with $\theta = 0$ are found, it is natural to try to
cover the entire region of classically forbidden transitions of 
$E-N$ plane with
a deformation procedure,
by moving in small steps in $\theta$ and $T$.  The solution to the boundary 
value problem with $(T + \Delta T,\; \theta + \Delta \theta)$ may be
obtained numerically, by applying an iteration technique, with the known
solution at $(T,\;\theta)$ serving as the initial approximation\footnote{
In practice, the Newton--Raphson method is particularly convenient (see 
Refs.~\cite{Bonini:1999cn,Bonini:1999kj,Kuznetsov:1997az,Bezrukov:2001dg}).}.
Provided the solutions end up in correct asymptotic region at each step, 
i.e. $X\to +\infty$ on part  D of the contour, the 
solutions obtained by this  procedure of small deformations  are physically relevant. 
However, the method of small deformations fails to produce relevant solution 
 if there are bifurcation points in the $E$--$N$ plane, where the
physical branch of solutions  merges to an unphysical branch. As there are 
unphysical solutions close to physical ones in the vicinity of 
bifurcation points, one  cannot  use the procedure of small deformations
near these points.

\begin{figure}
  \begin{center}
    \includegraphics[width=\columnwidth]{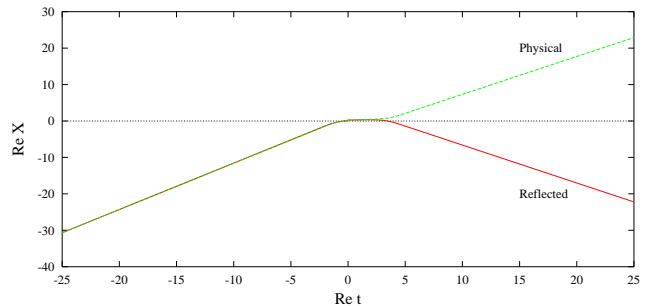}
  \end{center}
  \caption{The dependence of the tunneling coordinate $X$ on time for 
     two solutions with nearly the same energy and initial excitation number. 
	The physical solution tunnels to the asymptotic region $X\to +\infty$,
while the unphysical one gets reflected back to $X\to -\infty$.
      The physical solution has $E = 1.028,\;N =
    0.44$, while the unphysical one has $E = 1.034,\;N =
    0.44$. These two solutions are close to the point at the bifurcation
line $E_1(N = 0.44) = $ 1.031.}
  \label{fig9}
\end{figure}

We have found numerically that
in our model the method of small deformations produces correct solutions to the
$T/\theta$ boundary value problem in a large region of the $E-N$ plane, 
where $E < E_1(N)$.  However, at
sufficiently high energy $E > E_1(N)$, where
$E_1(N) \gtrsim E_\mathrm{S}$,  the deformation procedure 
generates solutions that bounce back from the barrier (see
Fig.~\ref{fig9}), i.e.\ have wrong ``topology''. 
This happens deep inside the region of classically forbidden  transitions,
where the suppression is large,
 and one naively expects the semiclassical technique to work well.
Clearly, solutions with wrong topology do not describe the tunneling transitions of
interest.  Therefore, if the semiclassical method is applicable at all
in the region $E_1(N) < E < E_0(N)$, there exists another, physical
branch of solutions.  In that case the line $E_1(N)$ is the
bifurcation line, where the physical solutions ``meet'' the ones with
wrong ``topology''.  Walking in small steps in $\theta$ and $T$ is
useless in the vicinity of this bifurcation line, and one needs to
introduce some trick to find the relevant solutions beyond that
line. The bifurcation line $E_1(N)$ for our quantum mechanical problem
of two degrees of freedom is shown in
Fig.~\ref{fig7}.

The loss of topology beyond a certain bifurcation line in $E-N$ plane
is by no means a property of our model only.  This phenomenon has been
observed in field theory models, in the context of both induced false
vacuum decay~\cite{Kuznetsov:1997az} and baryon-number violating
transitions in gauge theory~\cite{Bezrukov:2001dg} (in field-theoretic
models, the parameter $N$ is the number of incoming particles).  In
all cases, the loss of topology prevented one to compute the
semiclassical exponent for the transition probability in an
interesting region of relatively high energies.	

Coming back to quantum mechanics of two degrees of freedom, we
point out that the properties of tunneling solutions with different energies
approaching the
bifurcation line $E_1(N)$ from the left of the $E$--$N$ plane, 
are in some sense similar to
the properties of tunneling solutions in one--dimensional quantum
mechanics whose energy is close to the barrier height, 
see Appendix~\ref{sec:1D}. Again
by continuity, these solutions of our two-dimensional model spend a
long time in the interaction region; this time tends to infinity on
the line $E_1(N)$. Hence, at any point of this line, there is a
solution which starts in the asymptotic region left of the barrier, and
ends up on an excited sphaleron. Such a behavior is indeed possible because
of the existence of an unstable direction near the (excited)
sphaleron, even for complex initial data.  We suggest in the next
Section a trick to deal with this situation---this is our
regularization technique.

\section{\label{sec:4}Regularization technique}

In this Section we develop our regularization technique, and
find the physically relevant solutions between the lines $E_1(N)$ and
$E_0(N)$.  We will see that all solutions from the new branch (and not
only on the lines $E_0(N)$ and $E_1(N)$) correspond to tunneling onto
the excited sphaleron (``tunneling on top of the barrier'').  
These solutions would be very difficult, if at
all possible, to obtain directly, by solving numerically the
non-regularized classical boundary value problem~\eqref{bc}: they are
complex at finite times, and become real only asymptotically as $t\to
+\infty$, whereas numerical methods require working with finite time
intervals.

As an additional advantage, our
regularization technique enables one to obtain a family of the over--barrier
solutions, that covers all the region of the initial
data, corresponding to classicaly allowed transitions, including its 
boundary. This is  of interest in models with large number of degrees of 
freedom and in field theory, where finding the boundary $E_0(N)$ by direct 
methods is difficult (see e.g., 
Ref.~\cite{Rebbi:1995zw} for discussion of this point).

\subsection{\label{sec:reg_forb}Regularized problem: classically
forbidden transitions}

The main idea of our method  is
to regularize the equations of motion by adding a term
proportional to a small parameter $\epsilon$ so that configurations staying 
for an infinite time near the sphaleron no
longer exist among the solutions of the $T/\theta$ boundary value
problem.  After performing the regularization we explore all
region of classically forbidden transitions without crossing 
the bifurcation line.
Taking then the limit ${\epsilon \to 0}$ we reconstruct the correct
values of $F$, $E$ and $N$.

When formulating the regularization technique it is more convenient to
work with the functional $F[X,y;X^*,y^*;T,\theta]$,~Eq.~\eqref{F}, itself rather than
with the equations of motion. We prevent $F$ from beeing
extremized by configurations
approaching the excited sphalerons asymptotically.  To achieve this,
we add to the original functional~\eqref{F} a new term of the form $2
\epsilon T_{\mathrm{int}}$, where $T_{\mathrm{int}}$ estimates the
time the solution ``spends'' in the interaction region.  The parameter
of regularization $\epsilon$ is the smallest one in the problem, so
any regular extremum of the functional $F$ (the solution that spends
finite time in the region $U_{\mathrm{int}} \ne 0$) changes only
slightly after the regularization.  At the same time, the excited
sphaleron configuration has $T_{\mathrm{int}} = \infty$ which leads to
infinite value of the regularized functional $F_{\epsilon} \equiv F +
2 \epsilon T_\mathrm{int}$.  Hence, the excited sphalerons are not
stationary points of the regularized functional.

For the problem at hand $U_{\mathrm{int}} \sim 1$ in the interaction 
region, and $T_\mathrm{int}$ can be defined as follows,
\begin{equation}
T_{\mathrm{int}} = \frac12 \left[\int dt\; U_{\mathrm{int}}(X,y) + 
\int dt\; U_{\mathrm{int}} (X^*,y^*)\right] \;.\label{Tint}
\end{equation}
We notice that $T_\mathrm{int}$ is real, and that the regularization 
is equivalent to the 
multiplication of the interaction potential by a complex factor
\begin{equation}
U_{\mathrm{int}}\to (1 - i \epsilon) U_{\mathrm{int}} = 
\mathrm{e}^{- i \epsilon} U_{\mathrm{int}}  + O(\epsilon^2)
\;.\label{Uintnew}
\end{equation}
This results in the corresponding change of the classical equations of
motion, while the boundary conditions~\eqref{bcplus},~\eqref{bcminus} 
remain unaltered.

We still have
to understand whether solutions with $\epsilon \ne 0$ exist at
all.  The reason for the existence of such solutions is as follows.
Let us consider a well-defined (for $\epsilon > 0$) matrix element
\begin{equation*}
  \mathcal{T}_{\epsilon} = \lim\limits_{t_f - t_i \to \infty}
  \sum\limits_{f} \left|\langle f|
\mathrm{e}^{ (- i \hat H - \epsilon U_\mathrm{int})(t_f - t_i)} 
| E,N \rangle\right|^2 \;,
\end{equation*}
where $|E,\;N\rangle$ denotes, as before, the incoming state with given energy 
and oscillator excitation number.  The quantity ${\cal T}_{\epsilon}$ has a well 
defined limit as $\epsilon \to 0$, equal to 
the tunneling probability~\eqref{TT}. As the saddle point of the regularized
functional $F_{\epsilon} \equiv F + 2 \epsilon T_{\mathrm{int}}$ gives
the semiclassical exponent for the quantity ${\cal
  T}_{\epsilon}$, one expects that such saddle point indeed exists.

Therefore, the regularized $T/\theta$ boundary value problem is
expected to have solutions necessarily spending finite time in the
interaction region. By continuity, these solutions do not experience 
reflection from the barrier,
if one makes use of the procedure of small deformations starting from solutions with correct ``topology''.
The line $E_1(N)$ is no longer a bifurcation line of the regularized 
system, so the procedure of small deformations enables one  to cover the entire
region of classically forbidden transitions.
The semiclassical  suppression factor of the original problem is recovered 
in the limit $\epsilon \to 0$.

It is worth noting that the interaction time is Legendre
conjugate to $\epsilon$,
\begin{equation}\label{TintThetaE}
  T_{\mathrm{int}} =  \frac{1}{2}
  \frac{\partial}{\partial \epsilon} F_{\epsilon}(E,N,\epsilon)
  \;.
\end{equation}
This equation  may be used as a check of numerical calculations. 

We implemented the regularization procedure numerically. 
To solve the boundary value problem, we make use of 
the computational methods described in Ref.~\cite{Bonini:1999cn,Bonini:1999kj}.
To obtain the semiclassical tunneling exponent in the region between
the bifurcation line $E_1(N)$ and the boundary of the region of classically
allowed transitions $E_0(N)$,
we began with a solution to the non-regularized problem deep in the 
``forbidden'' region of initial data 
(i.e., at $E < E_1(N)$). Then  we increased 
the value of $\epsilon$ from zero to a certain small positive number,
keeping $T$ and $\theta$ fixed.
Then we changed $T$ and $\theta$ in small steps, keeping $\epsilon$ finite,
and found solutions to the regularized problem in the region 
$E_1(N) < E < E_0(N)$. These solutions had correct ``topology'', i.e.\ they
indeed ended up in the asymptotic region $X \to +\infty$. Finally, we 
lowered $\epsilon$ and extrapolated $F$, $E$ and $N$ to the limit 
$\epsilon \to 0$.

\begin{figure}
\includegraphics[width=\columnwidth]{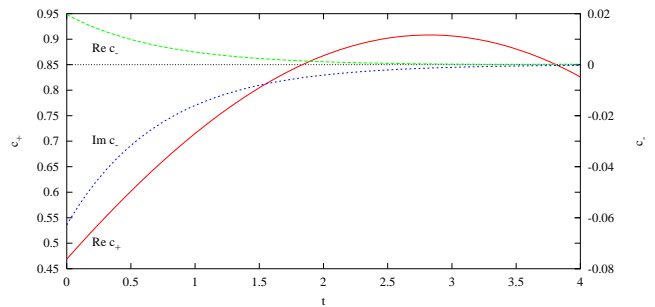}
\caption{The large-time behavior of a solution with 
$\epsilon = 0$ at $(E = 1.05,\; N=0.43)$.
The coordinates $X$ and $y$ are decomposed in the basis of the eigenmodes near 
the sphaleron.  Note that $\Imag c^+ = 0$.}
\label{fig14}
\end{figure}

Let us consider more carefully the solutions in the region $E_1(N) < E
< E_0(N)$ which we obtain in the limit $\epsilon\to0$.  They belong to
a new branch, and thus may exhibit new physical properties.  Indeed,
we found that, as the value of $\epsilon$ decreases to zero, the
solution at {\it any} point $(E,N)$ with $E_1(N) < E < E_0(N)$ spends
more and more time in the interaction region. The limiting solution
corresponding to $\epsilon = 0$ has infinite interaction time: in
other words, it tends, as $t\to +\infty$, to one of the excited
sphalerons.  The resulting physical picture is that at large enough
energy (i.e., at $E > E_1(N)$), the system prefers to tunnel exactly
onto an unstable classical solution, excited sphaleron, that
oscillates about the top of the potential barrier.  To demonstrate
this, we have plotted in Fig.~\ref{fig14} the solution
$\vec{x}(t)\equiv (X(t),y(t))$ at large times, after taking
numerically the limit $\epsilon \to 0$. To understand this figure, one
recalls that the potential near the sphaleron point $X = y = 0$ has
one positive mode and one negative mode.  Namely, by introducing new
coordinates $c_+$, $c_-$,
\begin{eqnarray}
&&X = \cos\alpha \;c_+ + \sin\alpha \; c_-\;,\nonumber\\
&&y = -\sin\alpha \;c_+ + \cos\alpha \; c_-\;,\nonumber\\
&& \mathrm{ctg } 2 \alpha = - \frac{\omega^2}{2}\;,\nonumber
\end{eqnarray}
one writes, in the vicinity of the sphaleron,
\begin{equation*}
H = 1 + \frac{p_+^2}{2} + \frac{p_-^2}{2} + \frac{\omega_+^2}{2} c_+^2 - 
\frac{\omega_-^2}{2} c_-^2\;,
\end{equation*}
where
\begin{equation*}
\omega_\pm^2 = \pm (- 1 + \frac{\omega^2}{2}) + \sqrt{1 + \frac{\omega^4}{4}}
> 0\;.
\end{equation*}
Since the solutions to the $T/\theta$ boundary value problem are
complex, the coordinates $c_+$ and $c_-$ are complex too.  We show in
Fig.~\ref{fig14} real and imaginary parts of $c_+$ and $c_-$ at large
real time $t$ (part CD of the contour).  We see that while $\Real c_+$
oscillates, the unstable coordinate $c_-$ approaches asymptotically
the sphaleron value: $c_- \to 0$ as $t \to +\infty$.  The imaginary
part of $c_-$ is non-zero at any finite time.  This is the reason for
the failure of straightforward numerical methods in the region $E >
E_1(N)$: the solutions from the physical branch do not satisfy the
conditions of reality at any large but finite final time.  We have
pointed out in Sec.~\ref{sec:Ttheta} that this can happen only if the
solution ends up near the sphaleron, which has a negative mode.
This is precisely what happens: for $\epsilon=0$ at asymptotically large
$t$ our solutions are real and oscillate near the sphaleron, remaining in the
interacton region.

\subsection{\label{sec:4.2}Regularization technique versus exact
    quantum--mechanical solution}

Quantum mechanics of two degrees of freedom is a convenient
testing ground for checking the semiclassical methods and, in particular, our
regularization technique. We have found the solutions to the full stationary
Schr\"odinger equation and exact tunneling probability ${\cal T}$ 
by applying the numerical technique of 
Refs.~\cite{Bonini:1999cn,Bonini:1999kj}.
Our numerical calculations were performed for several small
values of the semiclassical parameter $\lambda$, namely, for 
$\lambda = 0.01$---$0.1$. Transitions through the barrier for these
values of the semiclassical parameter are well suppressed. In particular, for 
$\lambda=0.02$ the tunneling probability ${\cal T}$ is of order 
$\mathrm{e}^{-14}$. To check the semiclassical result
with better precision, we have calculated the exact suppression exponent 
$F_\mathrm{QM}(\lambda) \equiv -\lambda\log{\cal T}$ (cf.~\eqref{+}) for
$\lambda=0.09, 0.05, 0.03, 0.02$ and extrapolated $F_{\mathrm{QM}}$ 
to $\lambda=0$ by polynomials of the third and fourth order. The results of 
extrapolation are independent of the order ($3$ or $4$) of polynomials 
with precision of $1\%$.
The extrapolated suppression exponent $F_{\mathrm{QM}} (0)$ 
corresponds to infinite suppression and must exactly coincide (up to numerical 
errors) with the correct semiclassical result. 

\begin{figure}
  \begin{center}
    \includegraphics[width=\columnwidth]{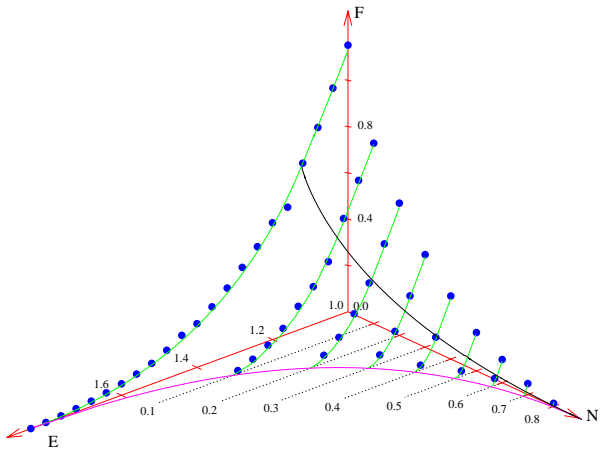}
  \end{center}
  \caption{The tunneling exponent $F(E,N)$
    in the region $E > E_{\mathrm{S}} = 1$.  The lines
    show the semiclassical results while points represent exact
    ones, obtained by solving the Schr\"odinger equation.  The
    lines across the plot are the 
    boundary of the region of classically allowed transitions  $E_0(N)$ and
    the bifurcation line $E_1(N)$.}
\label{fig16}
\end{figure}

We performed this check in the region $E > E_S = 1$, which is most 
interesting for our purposes. The results
of the full quantum mechanical calculation of the suppression exponent
$F_{\mathrm{QM}}$ in the limit $\lambda\to 0$ are represented by points in
Fig.~\ref{fig16}.  The lines in that figure represent the values
of the semiclassical exponent $F(E,N)$ for constant $N$, which we
obtain in the limit $\epsilon \to 0$ of the regularization procedure.  
In practice, instead of taking the limit $\epsilon \to 0$ one calculates the 
regularized functional $F_\epsilon (E,N) = F(E,N) + O(\epsilon)$ for
small enough $\epsilon$. In our calculations we used the value 
$\epsilon = 10^{-6}$, so that the value of the suppression exponent was
found with precision of order $10^{-5}$.
We see that in the entire region of classically forbidden transitions
(including the region $E > E_1(N)$) the semiclassical
result for $F$ coincides with the exact one.

\begin{figure}
  \begin{center}
    \includegraphics[width=\columnwidth]{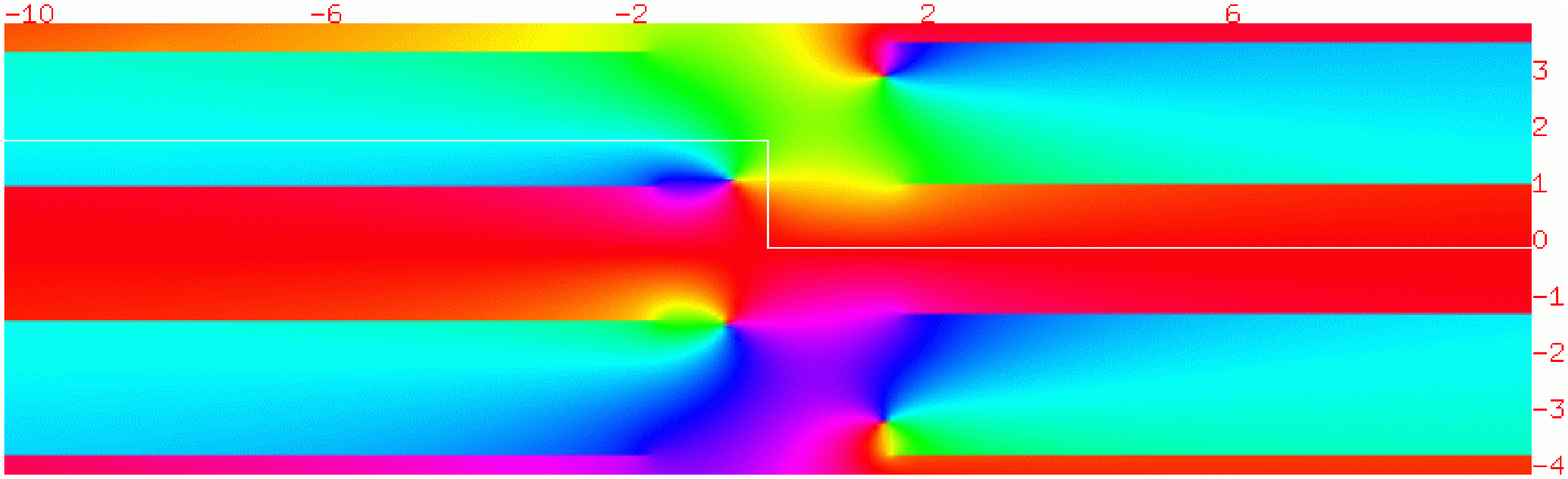}\\
    (a)\\
    \includegraphics[width=\columnwidth]{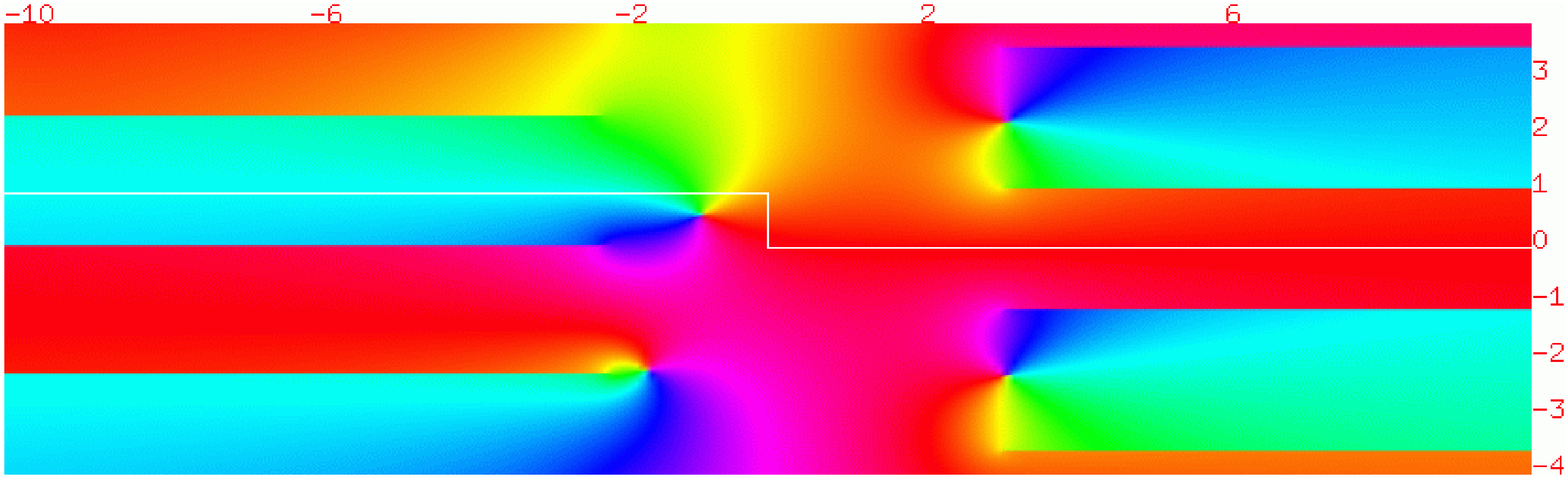}\\
    (b)\\
    \includegraphics[width=\columnwidth]{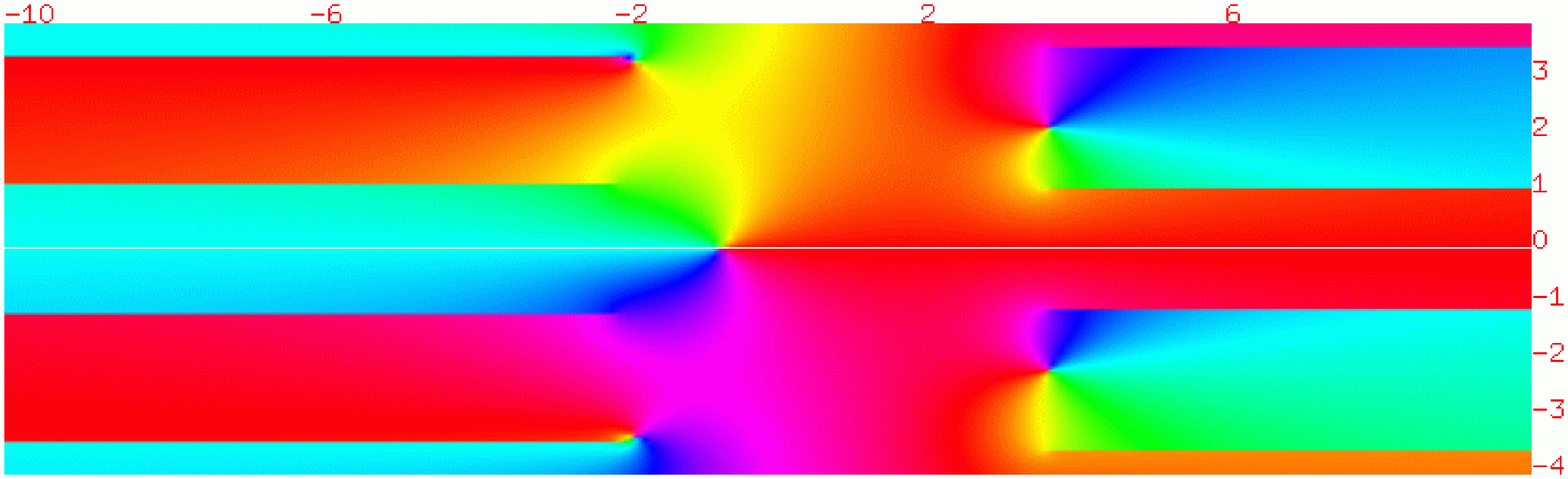}\\
    (c)\\
  \end{center}
  \begin{flushleft}
  $\begin{array}{c}\pi\\-\pi\end{array}
  \begin{array}{c}\includegraphics[width=1cm]{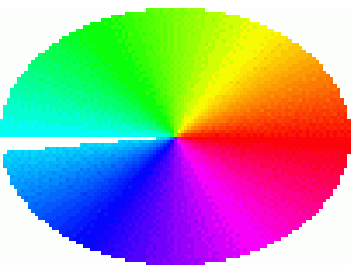}\end{array}
  0$
  \end{flushleft}
  \caption{The phase of the tunneling coordinate in complex time 
    plane at the three points of the curve $\tau = 380$,
    $\vartheta = 130$. Points (a), (b), (c) correspond to 
    $\epsilon = \epsilon_a=0.01$, 
    $\epsilon = \epsilon_b=0.0048$ and $\epsilon = \epsilon_c=0$, 
    respectively. The 
    asymptotics $X\to -\infty$ and $X\to +\infty$ correspond to
    $\mathrm{arg}(X) = \pi$ and $0$.  The contour in
    the time plane is plotted with white line.}
\label{fig18}
\end{figure}


\subsection{\label{sec:4.3}Classically allowed transitions}

Let us show that our regularization procedure enables one to
obtain a subset of classical over--barrier solutions, existing at
high enough energies. This subset is interesting, as it extends all
the way to the boundary of the region of classically allowed 
transitions, $E =E_0(N)$. 
In principle, finding this boundary is purely a problem of
classical mechanics, and, indeed, in mechanics of two degrees of
freedom one obtains this boundary numerically by solving 
the Cauchy problem for given $E$ and $N$ and all possible oscillator
phases, see Sec.~\ref{sec:over-barrier}.  However, if the number of
degrees of freedom is much larger, this classical problem becomes
quite complicated, as one has to span a high-dimensional space of
Cauchy data.  As an example, a stochastic Monte Carlo technique was
developed in Ref.~\cite{Rebbi:1995zw} to deal with this problem in
field theory context.  The approach below is as an
alternative to the Cauchy methods.

First, let us recall that all classical over--barrier solutions with
given energy and excitation number satisfy the $T/\theta$ boundary
value problem with $T = 0$, $\theta = 0$.  We cannot reach the ``allowed''
region of $E-N$ plane without regularization, since  we have to cross
the line $E_0(N)$ corresponding to the excited sphaleron
configurations in the final state.  However, the excited sphalerons are 
longer exist among the solutions of the regularized boundary value
problem at any finite value of $\epsilon$.  This suggests that the
regularization enables one to enter the region of the classically allowed
transitions
and, after taking an appropriate limit, obtain classical solutions
with finite values of $E$, $N$.

By definition, the classically allowed transitions have $F = 0$.
Thus, one expects that in the ``allowed'' region of initial data, 
the regularized problem
has the property that 
$F_{\epsilon}(E,N) = \epsilon f(E,N) +
O(\epsilon^2)$.  In view of the inverse Legendre formulas 
(\ref{T}),~(\ref{theta}) the values of $T$ and $\theta$ must be of 
order $\epsilon$: $T
= \epsilon\tau(E,N)$, ${\theta = \epsilon
  \vartheta(E,N)}$, where the quantities $\tau$ and
$\vartheta$ are related to the initial energy and excitation number
(see Eqs.~(\ref{T}),~(\ref{theta})) in the following way,
\begin{eqnarray}
  \tau &=& -\lim\limits_{\epsilon \to 0} \frac{\partial}{\partial E} 
    \frac{F_{\epsilon}}{\epsilon} = 
  -\frac{1}{2}\frac{\partial}{\partial E}  T_\mathrm{int}(E,N)
  \label{Ecl}
  \;,\\
  \vartheta &=& -\lim\limits_{\epsilon \to 0} \frac{\partial}{\partial N} 
    \frac{F_{\epsilon}}{\epsilon} = 
  -\frac{1}{2}\frac{\partial}{\partial N}  T_\mathrm{int}(E,N)\;,
  \label{Ncl}
\end{eqnarray}
where we have used Eq.~\eqref{TintThetaE}.
Therefore, one expects that one
can invade the region of classically allowed transitions
by taking a fairly sophisticated limit
$\epsilon \to 0$ with  $\tau \equiv T/\epsilon = \mathrm{const}$,
$\vartheta\equiv \theta/\epsilon = \mathrm{const}$.
For the allowed transitions the parameters $\tau$ and $\vartheta$ are 
analogous to $T$ and $\theta$.  

By solving the regularized $T/\theta$ boundary 
value problem one
constructs a single solution for given $E$ and $N$.  
On the other hand, for $\epsilon = 0$ there are more classical
over--barrier solutions: they form a  continuous family labeled by  
the initial oscillator phase.
Thus, after taking the limit $\epsilon \to 0$ one obtains a subset of
 over--barrier solutions,  which should therefore obey
some additional constraint.  It is almost obvious, that this constraint is 
that the interaction time $T_{\mathrm{int}}$, Eq.~(\ref{Tint}), is minimal.  
This is shown in Appendix~\ref{app:B}.

The  subset of classical over--barrier solutions obtained in
the limit ${\epsilon \to 0}$ of the regularized $T/\theta$ procedure
extends all the way to the boundary of the region of classically allowed
transitions.
Let us see what happens as one approaches this boundary from the ``classically 
allowed'' side. At the boundary $E_0(N)$, the unregularized solutions tend to 
excited sphalerons, so the interaction time $T_\mathrm{int}$ is infinite.
This is consistent with~(\ref{Ecl}),~(\ref{Ncl})  only if $\tau$ and 
$\vartheta$ become infinite at the boundary. 
Thus, to obtain a point of the boundary one takes the further limit,
\begin{equation*}
  \big( E_0(N),\, N \big) =
  \lim_{\stackrel{\scriptstyle \tau/\vartheta = \mathrm{const}}
                 {\scriptstyle \tau \to +\infty}}
  \big( E(\tau,\vartheta), N(\tau,\vartheta) \big) \;.
\end{equation*}
Different values of $\tau/\vartheta$ correspond to different points of the
line $E_0(N)$.
In this way one finds  the boundary of the region of classically
allowed transitions without an initial-state simulation.  

We have checked this procedure numerically.   The limit $\epsilon \to 0$ exists
indeed---the values of $E$ and $N$ tend to the  point  
of the $E$--$N$ plane, which corresponds to the classically allowed transition.
The phase of the tunneling coordinate
$X(t)$ in complex time plane is shown in Fig.~\ref{fig18} for the
three points (a), (b) and (c) of the curve 
$\tau \equiv T/\epsilon = 380,\;\vartheta\equiv \theta/\epsilon = 130$. 
Point (a) lies deep inside the tunneling region, $E_a < E_1(N_a)$,
point (c) corresponds to over-barrier solution with 
$T=0,\;\theta=0,\;\epsilon=0$, point (b) is in the middle of the curve. The
branch points of the solution\footnote{The phase of the tunneling
coordinate turns by $\pi$ around the branch point.  The points where
the phase of the tunneling coordinate turns by $2\pi$ correspond to
the zeroes of $X(t)$.}, the cuts and the contour are clearly seen on
these graphs.

It is worth noting that the left branch points move down as
$T$ and $\theta$ approach zero.  Solutions close enough to the 
boundary $E_0(N)$ have left branch point in
the lower complex half-plane, see Fig.~\ref{fig18}.  Therefore, the
corresponding contour may be continuously deformed to the real time
axis.  These solutions still satisfy the reality conditions
asymptotically (see Fig.~\ref{fig14}), but show nontrivial complex
behavior at any finite time.

\begin{figure}
  \begin{center}
    \includegraphics[width=0.9\columnwidth]{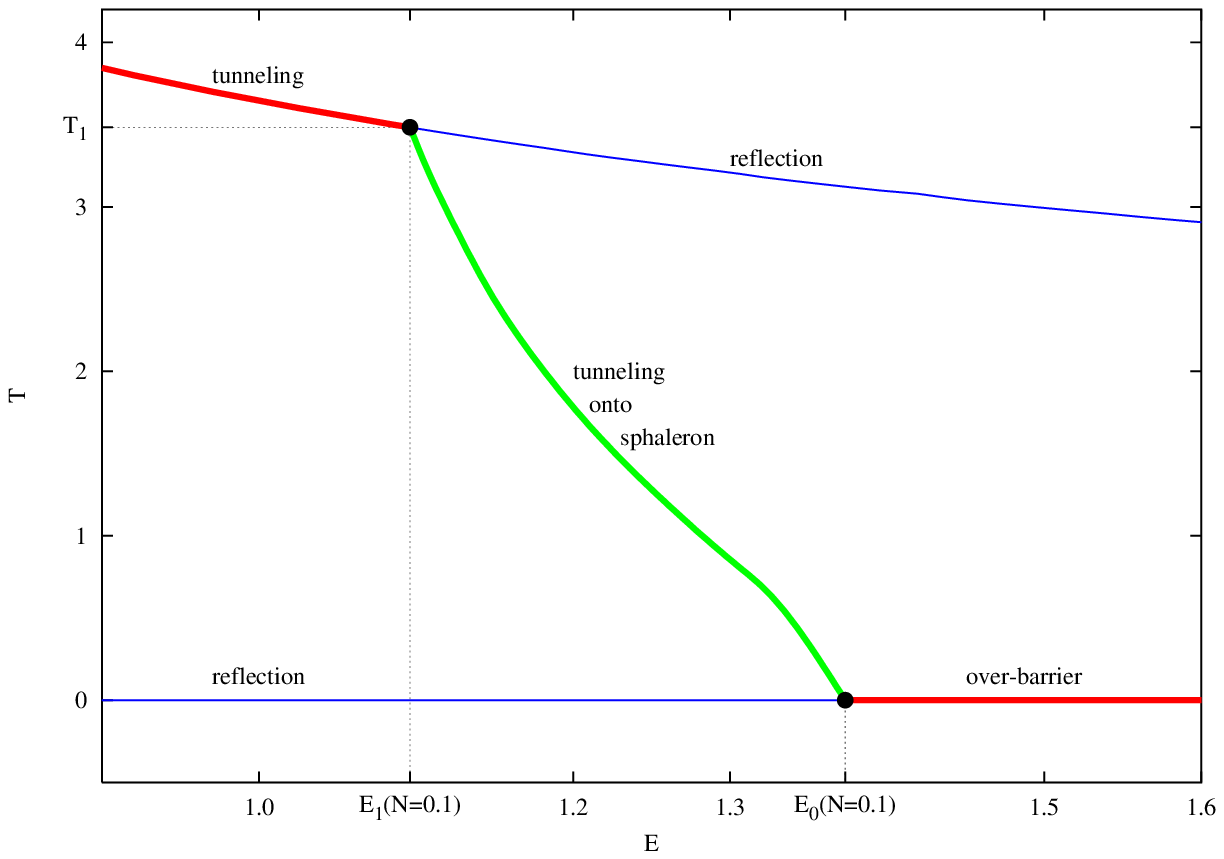}\\
	(a)\\
    \includegraphics[width=0.9\columnwidth]{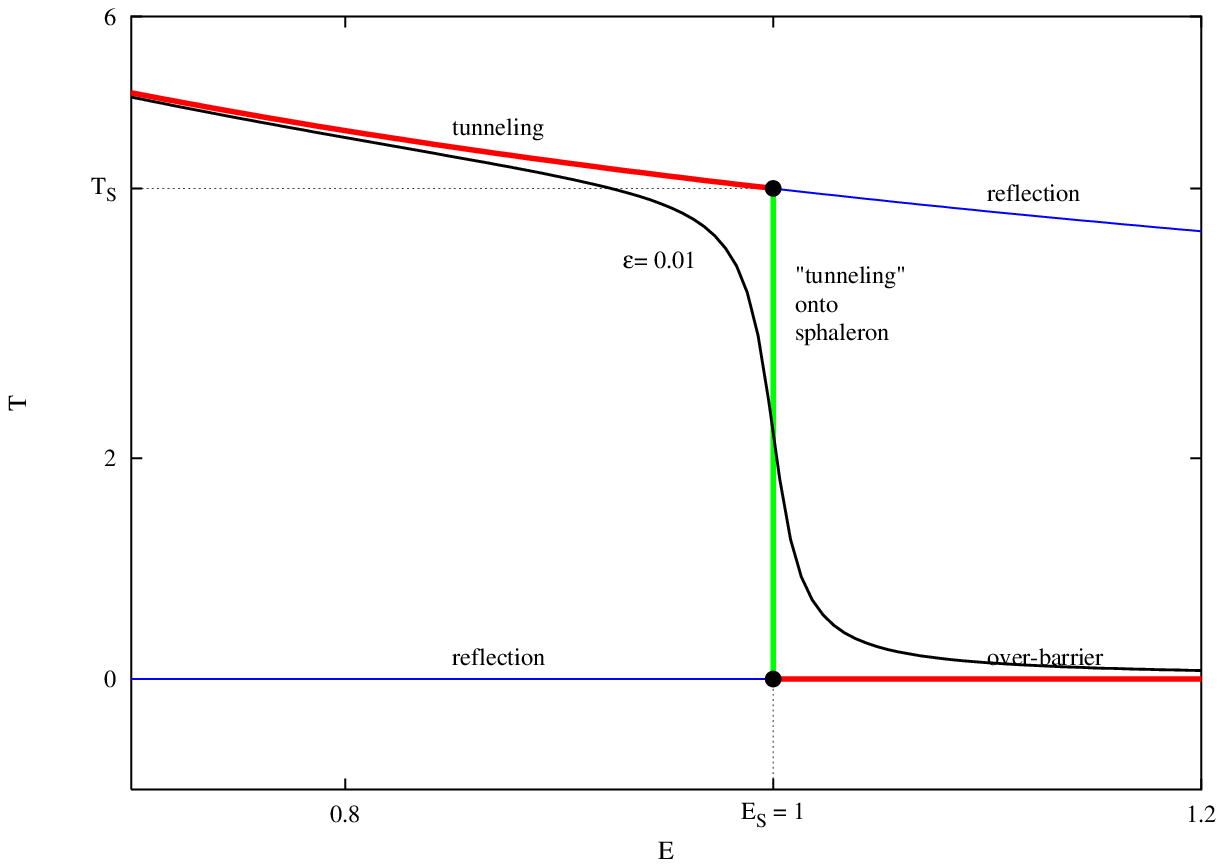}\\
	(b)
  \end{center}
  \caption{Dependence of the parameter $T = - \partial F/\partial E$ on energy 
for (a) two-dimensional model with fixed $N=0.1$ and (b) one-dimensional model
(see Appendix~\ref{sec:1D}).
Different lines correspond to different branches of classical solutions
to $T/\theta$ boundary value problem. The branches labelled ``reflection''
end up on the wrong side of the barrier. Figure (b) contains also a line 
with nonzero~$\epsilon$.}
\label{fig15}
\end{figure}

Making use of the regularized $T/\theta$ procedure, one is able to
approach the boundary of the region of classically allowed transitions
from both
sides.  The points at this boundary are obtained by taking the
limits $T \to 0,\; T/\theta = \mathrm{const}$ of the tunneling
solutions and $\tau \to +\infty$, ${\tau/\vartheta = \mathrm{const}}$
of the classically allowed ones.  As $\tau^* \equiv \tau/\vartheta =
T/\theta$ by construction, the lines $\tau^* = \mathrm{const}$ are
continuous at the boundary $E_0(N)$, though may have discontinuity of
the derivatives.  The variable $\tau^*$ can be used to parametrize the curve
$E_0(N)$.

\section{Conclusions}

We conclude that classical solutions describing transmissions of a bound 
system through a potential barrier with different values of energy
and initial oscillator excitation number form three branches.
These branches  merge at bifurcation lines $E_0(N)$ and $E_1(N)$.
Solutions from different branches describe 
physically different transition processes. Namely, solutions
at low energies
$E < E_1(N)$ describe conventional potential--like 
tunneling, while
at $E > E_0(N)$ they correspond to unsuppressed over-barrier transitions.
At intermediate energies, $E_1(N) < E < E_0(N)$, physically relevant solutions
describe transitions on top of the barrier.
This branch structure is shown in Fig.~\ref{fig15}a, where the period 
$T = \partial F/\partial E$ obtained numerically for the solutions from 
different branches is plotted as function of energy 
for $N=0.1$.

One notices that the qualitative structure of branches in model, with 
internal degrees of freedom is similar to the structure of branches in 
one-dimensional quantum mechanics (see Appendix~\ref{sec:1D}).
The latter is shown in Fig.~\ref{fig15}b. The features of  solutions 
in both cases are similar, though the solutions ending up on top of 
the barrier are degenerate in energy in one-dimensional case, and hence
are not really physically interesting.

In this paper we introduced the regularization technique which enables 
one to connect smoothly solutions from different branches. Its advantage 
is that it automatically chooses the physically 
relevant branch. This technique is particularly convenient in numerical 
studies: we have seen that it enables one to cover the whole 
interesting region of parameter space. We  applied this
technique to baryon number violating processes in electroweak 
theory\cite{Bezrukov:2003er}.

\begin{acknowledgments}
The authors are indebted to V.~Rubakov and C.~Rebbi for numerous
valuable discussions and criticism, A.~Kuznetsov, W.Miller 
and S.~Sibiryakov for
helpful discussions, and S.~Dubovsky, D.~Gorbunov, A.~Penin and
P.~Tinyakov for stimulating interest.  We wish to thank Boston
University's Center for Computational Science and Office of
Information Technology for allocations of supercomputer time.  This
reseach was supported by Russian Foundation for Basic Research grant
02-02-17398, grant of the President of the Russian Federation  
NS-2184.2003.2,  U.S.  Civilian Research and Development Foundation for
Independent States of FSU~(CRDF) award RP1-2364-MO-02, and under DOE
grant US DE-FG02-91ER40676. F.B. work is supported by the Swiss Science 
Foundation grant~7SUPJ062239.
\end{acknowledgments}

\appendix

\section{\label{app:A}$T/\theta$ boundary value problem}

The semiclassical method for calculating the probability of
tunneling from a state with a few parameters fixed was developed in
\cite{Rubakov:1992ec,Rubakov:1992fb, Kuznetsov:1997az,Bezrukov:2001dg} 
in context of field theoretical models and in
\cite{Miller:1970,Miller:1972,George:1972,Bonini:1999cn,Bonini:1999kj} 
in quantum mechanics.  Here
we outline the method, adapted to our model of two degrees of freedom.

\subsection{\label{app:A.1}Path integral representation of the
transition probability}

We begin with the path integral representation for the probability 
of tunneling  from the asymptotic region 
$X\to -\infty$ through a potential barrier.
Let the incoming state $|E,\;N\rangle$ have fixed energy and oscillator 
excitation number, and has support only for $X\ll 0$, well outside the range 
of the potential barrier. 
The inclusive tunneling probability for states of this type is given by
\begin{multline}
\mathcal{T}(E,N) 
= \lim\limits_{t_f - t_i \to \infty} \Bigg\{ \int\limits_0^{+\infty}
dX_f \int\limits_{-\infty}^{+\infty} dy_f \\
\left|\langle X_f,y_f|\mathrm{e}^{-i\hat{H}(t_f - t_i)}|E,N\rangle
\right|^2\Bigg\}\;,
\end{multline}
where $\hat{H}$ is the Hamiltonian operator.
This probability can be reexpressed in terms of the
transition amplitudes 
\begin{equation}
{\cal A}_{fi} = \langle X_f,\;y_f| \mathrm{e}^{-i \hat{H}(t_f - t_i)} |X_i,\;y_i\rangle
\end{equation}
and initial-state matrix elements
\begin{equation}
{\cal B}_{ii'} = \langle X_i,\;y_i| E,\;N\rangle
\langle E,\;N| X_i',\;y_i'\rangle
\end{equation}
in the following way,
\begin{multline}\label{Ttemp}
{\cal T}(E,\;N) = \lim\limits_{t_f-t_i \to \infty}\Bigg\{
\int\limits_{0}^{+\infty} dX_f \int\limits_{-\infty}^0
d{X}_i\,d{X}_i'\\
\int\limits_{-\infty}^{+\infty} dy_i\,dy_i'\,dy_f\;
 {\cal A}_{fi} 
{\cal A}_{i'f}^*
{\cal B}_{ii'}\Bigg\}\;.
\end{multline}
The transition amplitude and its complex conjugate have the familiar path integral 
representation:
\begin{eqnarray}\label{A}
  \mathcal{A}_{fi} &=& \int [d\vec{x}]
  \Bigg|_{\stackrel{
    \scriptscriptstyle\vec{x}(t_i) = \vec x_i}{
    \scriptscriptstyle\vec{x}(t_f) = \vec x_f}}
  \e^{iS[\vec{x}]}
  \;,\\
  \mathcal{A}_{i'f}^* &=& \int [d\vec{x'}]\Bigg|_{\stackrel{
    \scriptscriptstyle\vec{x}'(t_i) = \vec x'_i}{
    \scriptscriptstyle\vec{x}'(t_f) = \vec x_f}}
  \e^{-iS[\vec{x'}]}
  \;,\nonumber
\end{eqnarray}
where $\vec{x} = (X,y)$, and  $S$ is the action of the model.  To obtain a similar  
representation for the initial-state matrix elements, let 
us rewrite ${\cal B}_{ii'}$ as follows,
\begin{equation}
{\cal B}_{ii'}  = 
\langle X_i,\;y_i| \hat{P}_E \hat{P}_N|X_i',\;y_i'\rangle\;,
\end{equation}
where $\hat{P}_N$ and $\hat{P}_E$ denote the projectors onto 
states with oscillator excitation number $N$ and total energy 
$E$ respectively.  
It is convenient to use the coherent state formalism for the 
$y$-oscillator and choose the momentum basis for the $X$-coordinate.
In this representation, the 
kernel of the projector operator $\hat{P}_E\hat{P}_N$ takes the form
\begin{multline*}
\langle q,\;b| \hat{P}_E \hat{P}_N | p,\;a\rangle = \frac{1}{(2 \pi)^2}
\int d\xi\,d \eta\\ 
\mathrm{exp}\left(- iE\xi  - iN\eta + \frac{i}2 p^2\xi + 
\mathrm{e}^{i \omega \xi + i\eta} \bar{b} a\right)\delta(q - p)\;,
\end{multline*}
where $|p,\;a\rangle$ is the eigenstate of the 
center-of-mass momentum $\hat{p}_X$ and $y$-oscillator annihilation 
operator $\hat{a}$ with eigenvalues $p$ and $a$ 
respectively.
It is straightforward to express this matrix element in the 
coordinate representation using the formulas
\begin{eqnarray*}
\langle y | a \rangle  &=&  \sqrt[4]{\frac\omega\pi} 
\mathrm{e}^{ - \frac12 a^2 + 
\sqrt{2 \omega} a y - \frac12 \omega y^2}\;,\\
\langle X | p \rangle  &=&  \frac{1}{\sqrt{2 \pi}} \mathrm{e}^{i p X}\;.
\end{eqnarray*}
Evaluating  the Gaussian integrals over $a$, $b$, $p$, $q$, we obtain
\begin{eqnarray}
\label{B}
{\cal B}_{ii'} &=& \int d\xi \; d\eta\;  \mathrm{exp}\left(
- i E \xi - i N \eta - \frac{i}{2} \frac{(X_i - X_i')^2}{\xi} \right.\nonumber\\
&&+\left.\frac{\omega}{1 - \mathrm{e}^{- 2 i \omega\xi - 2 i \eta}}\left[
\frac{y_i^2 + y_i^{\prime 2}}{2}(1 + \mathrm{e}^{- 2 i \omega \xi - 2 i \eta})
 \right.\right.\nonumber\\
&&\quad\quad\quad\quad\quad\quad\quad\quad\quad -2 y_i y_i' \mathrm{e}^{- i \omega \xi - i \eta}\bigg]\bigg)
\end{eqnarray}
where we omit the pre-exponential factor depending on $\eta,\;\xi$.
For the subsequent formulation of the boundary value problem it is convenient 
to introduce the notations
\begin{equation*}
  T = - i \xi\;,\qquad  \theta = - i \eta\;.
\end{equation*}
Then, combining together the integral representations (\ref{B}) and~(\ref{A})
and rescaling coordinates, energy and excitation number $\vec{x} \to 
\vec{x}/\sqrt{\lambda}$, $E\to E/\lambda$, $N\to N/\lambda$, we finally
obtain
\begin{multline}
\label{TPI}
{\cal T}(E,\;N) = \lim\limits_{t_f-t_i \to \infty} \Bigg\{
\int\limits_{-i\infty}^{+i\infty}  dT\;d\theta \int \;
[d\vec{x}\;d\vec{x}'] \\
\exp\left\{- \frac{1}{\lambda}F[\vec{x},\;\vec{x}';\;T,\;\theta]\right\}\Bigg\}\;,
\end{multline}
where 
\begin{multline}
F[\vec{x},\;\vec{x}';\;T,\;\theta] = - i S[X,y] + i S[X',\;y'] \\
- E T - N \theta + B_i(\vec{x}_i, \vec{x'}_i;
T,\theta).\label{FF}
\end{multline}
Here the non-trivial initial term $B_i$ is 
\begin{align}\label{Bi}
  B_i = \Bigg\{ & \frac{(X_i - X_i')^2}{2 T}
  \nonumber \\
    & - \frac{\omega}{1 - \mathrm{e}^{2 \omega T + 2 \theta}}\bigg[
      \frac12(y_i^2 + {y_i'}^2)(1 + \mathrm{e}^{2 \omega T +  2 \theta})
  \nonumber\\
    & \hphantom{- \frac{\omega}{1 - \mathrm{e}^{2 \omega T + 2 \theta}}\bigg[}
      - 2 y_i y_i' \e^{\omega T + \theta}\bigg]\Bigg\}.
\end{align}
In~\eqref{TPI} $\vec{x}$ and $\vec{x'}$ are independent integration
variables, while $\vec{x}_f'\equiv \vec{x}_f$, see Eqs.~\eqref{Ttemp}.

\subsection{\label{app:A.2}The boundary value problem}

For  small $\lambda$, the path integral \eqref{TPI}
is dominated by a stationary point of the functional $F$. 
Thus, to calculate the tunneling probability exponent, we extremize this 
functional with respect to all variables of integration: $X(t)$, $y(t)$, 
$X'(t)$, $y'(t)$, $T$, $\theta$. Note that because of the limit $t_f - t_i 
\to + \infty$, the  variation 
with respect to the initial and final values of coordinates leads to 
boundary conditions imposed at \emph{asymptotic} $t\to \pm\infty$, rather 
than at finite times $t_i,t_f$.
Note also that the  stationary points may be complex.

The variation of the functional \eqref{FF} with respect to the coordinates 
at intermediate times gives second order equations 
of motion, in general complexified,
\begin{subequations}
\label{problem1}
\begin{equation}
\frac{\delta S}{\delta X(t)} = 
\frac{\delta S}{\delta y(t)} = 
\frac{\delta S'}{\delta X'(t)} = 
\frac{\delta S'}{\delta y'(t)} = 0 \;. \label{equation}
\end{equation}
The boundary conditions at the final time $t_f \to +\infty$ are obtained by 
extremization 
of $F$ with respect to $X_f\equiv X_f'$, $y_f\equiv y_f'$.
These are 
\begin{equation}\label{bcf}
  \dot{X_f} = \dot{X_f'}
  \;,\qquad
  \dot{y_f} = \dot{y_f'}
  \;.
\end{equation}
It is convenient to write the conditions at the initial time (obtained 
by varying $X_i$, $y_i$, $X_i'$, $y_i'$) in terms of the asymptotic quantities.
At the initial moment of time $t_i \to -\infty$, the system moves in the region $X\to -\infty$,
well outside of the range of the potential barrier.  
Equations \eqref{equation} in this region describe free motion of decoupled oscillator,
and the general solution takes the following form,
\begin{eqnarray}
&&X(t) = X_i + p_i(t - t_i)\nonumber,\\
&&y(t) = \frac{1}{\sqrt{2 \omega}} \left[a \mathrm{e}^{ - i \omega(t - t_i)}+
\bar{a} \mathrm{e}^{i \omega(t - t_i)}\right]\nonumber\;,
\end{eqnarray}
while the solution for $X'(t)$, $y'(t)$ has similar form.
For the moment, $a$ and $\bar{a}$ are independent variables.
The initial boundary conditions in terms of the asymptotic variables 
$X_i$, $p_i$, $a$, $\bar{a}$ take the form:
\begin{eqnarray}
&&p_i = p_i' = - \frac{X_i - X_i'}{i T},\nonumber\\
&&a'+\bar{a}' = a\mathrm{e}^{\omega T + \theta} + \bar{a} \mathrm{e}^{- \omega T
- \theta} \label{bci}\;,\\
&&a+\bar{a} = a'\mathrm{e}^{-\omega T - \theta} + \bar{a}' \mathrm{e}^{\omega T
+ \theta} \nonumber\;.
\end{eqnarray}
The variation with respect to the Lagrange multipliers $T$ and $\theta$ 
gives the relation between the values of $E$, $N$ and initial asymptotic 
variables (here we use the boundary conditions \eqref{bci}),
\begin{eqnarray}
  E &=& \frac{p_i^2}{2} + \omega N\label{EN0}\;,\\
  N &=& a\bar{a}\nonumber\;.
\end{eqnarray}
Equations \eqref{equation} -- \eqref{EN0} constitute the complete set of 
saddle-point equations for the functional $F$. 

The variables $X'$ and $y'$ originate from the conjugate
amplitude ${\cal A}_{i'f}^*$ (see Eq.~\eqref{A}), which suggests that
they are the complex conjugate to $X$, $y$. Indeed, the Ansatz ${X'(t) =
X^*(t)}$, ${y'(t) = y^*(t)}$ is compatible with the boundary value
problem \eqref{problem1}. Then the Lagrange multipliers $T$,
$\theta$ are real, and the problem \eqref{problem1} may
be conveniently formulated at the contour ABCD in the complex time
plane (see Fig.~\ref{fig2}).
\end{subequations} 

Now we have only two independent complex variables $X(t)$ and $y(t)$,  
that have to satisfy the classical equations of motion in the interior of the 
contour,
\begin{subequations}
\label{problem2}
\begin{equation}
\frac{\delta S}{\delta X(t)} = 
\frac{\delta S}{\delta y(t)} =0\;. \label{equation2}
\end{equation}
The final boundary conditions (see Eq.~\eqref{bcf}) become the
conditions of the reality of the variables $X(t)$ and $y(t)$ at the
asymptotic part D of the contour:
\begin{eqnarray}
\begin{array}{l}
\Imag X_f = 0,\;\; \Imag y_f = 0\;,\\[1ex] 
\Imag \dot{X}_f = 0\;\;\Imag\dot{y}_f = 0\;,
\end{array} \qquad t\to +\infty\label{bcf2}\;.
\end{eqnarray}
The seemingly  complicated initial conditions \eqref{bci} simplify when written 
in terms of the  time coordinate $t' = t + i T/2$ running along the part AB of 
the contour. Let us again write the asymptotics of a solution, but now 
along the initial part AB of the contour:
\begin{eqnarray}
&&X = X_0 + p_0 (t' - t_i)\;,\nonumber \\ 
&&y = \frac{1}{\sqrt{2 \omega}}\left[u \mathrm{e}^{ - i \omega(t' - t_i)} +
v\mathrm{e}^{i \omega(t' - t_i)}\right]\;.\nonumber
\end{eqnarray}
In terms of $X_0$, $y_0$, $u$ and $v$, the boundary conditions \eqref{bci} are
\begin{eqnarray}
&&\Imag X_0 = 0,\;\; \Imag p_0 = 0\;,\label{bci2}\\
&&v = u^*\mathrm{e}^\theta\;.\nonumber
\end{eqnarray}
\end{subequations} 
Finally, we write Eqs.~\eqref{EN0} in terms 
of the asymptotic variables along the initial part of the contour:
\begin{eqnarray}
&&E = \frac{p_0^2}{2} + \omega N\;,\label{EN}\\
&&N = \omega u v\;.\nonumber
\end{eqnarray}
These equations determine 
the Lagrange multipliers $T,\;\theta$ in terms of $E$, $N$. Alternatively,
we can solve the problem \eqref{problem2} for given values of $T$, $\theta$
and find the values of $E$,$N$ from Eqs.~\eqref{EN}, what is 
more convenient computationally.

Given a solution to the problem \eqref{problem2}, the exponent $F$ 
is the value of the functional~\eqref{FF} at this saddle point. In this way
we obtain the expression \eqref{Fshort} for the tunneling exponent.
The exponent $F$ is expressed now in terms of $S_0$, eq.~\eqref{8*} ---
integrated by parts action of the system. Non--trivial boundary term $B_i$,
eq.~\eqref{Bi}, is cancelled by the boundary term coming from 
the integration by parts.
Note that we did not make use of the constraints~\eqref{EN} to obtain
the formula \eqref{Fshort}, so we still have to extremize \eqref{Fshort} with
respect to $T$ and $\theta$ (see discussion in Sec.~\ref{sec:Ttheta}).

The classical problem \eqref{problem2} is conveniently dubbed
$T/\theta$ boundary value problem.  Eqs.~\eqref{bcf2} and \eqref{bci2}
imply eight real boundary conditions for two complex second-order
differential equations \eqref{equation2}.  However, one of these real
conditions is redundant: Eq.~\eqref{bcf2} implies that the (conserved)
energy is real, so the condition $\Imag p_0\to0$ is automatically
satisfied (note that the oscillator energy $E_\mathrm{osc}=\omega
uv=\omega\e^\theta uu^*$ is real).  On the other hand, the system
\eqref{problem2} is invariant under time translations along the real
axis.  This invariance is fixed, e.g., by demanding that
$\Real(X)$ takes a prescribed value at a prescribed large negative
time $t'_0$ (note that other ways may be used instead.  In particular,
for $E<E_1(N)$ it is convenient to impose the constraint
$\Real\dot{X}({t=0})=0$).  Together with the latter requirement, we
have exactly eight real boundary conditions for the system of two
complexified (i.e.\ four real) second-order equations.

\section{\label{app:B}A property of solutions to $T/\theta$ problem in
for the case of over--barrier transitions}

For given $E$, $N$ there is only one over--barrier classical solution
which is  
obtained in the limit $\epsilon \to 0$ of the regularized $T/\theta$ 
procedure. To see what singles out this solution,  let us analyze the regularized functional
\begin{equation}
F_{\epsilon}[q] = F[q] + 2 \epsilon T_{\mathrm{int}}[q],
\end{equation}
where $q$ denotes the variables $\vec{x}(t),\; \vec{x}'(t)$ and 
$T,\;\theta$ together.  
The unregularized functional $F$ has a
valley of extrema $q^{e}(\varphi)$ 
corresponding to different values of the initial oscillator phase $\varphi$.
Clearly, at small $\epsilon$ the extremum of $F_\epsilon$ is close to a point 
in this valley with the phase extremizing $T_\mathrm{int}[q^e(\varphi)]$.
\begin{equation}
  \frac{d}{d\varphi} T_{\mathrm{int}}[q^e(\varphi)] = 0\;.
\end{equation}
Hence, the solution  $q^e_{\epsilon}$ 
of the regularized $T/\theta$ boundary value problem 
tends to the over--barrier classical solution, 
with $T_\mathrm{int}$ extremized with respect to 
the initial oscillator phase.  

Because $U_{\mathrm{int}}(\vec{x}) > 0$, $T_{\mathrm{int}}$ is a positive 
quantity with
at least one minimum.  In normal 
situation there is only one saddle point of $F_\epsilon$, so by solving 
the  $T/\theta$ boundary value problem one obtains the classical solution
with the time of interaction minimized.

\section{\label{sec:1D}Classically allowed transitions:
 one-dimensional example}

The difficulties with bifurcations 
of classical solutions emerge in quite a general class of 
quantum mechanical models.
To illustrate this statement,
let us consider the case of one-dimensional quantum mechanics,
where the result is given by well-known WKB formula.
We will show that the origin of the above 
difficulties can be seen in one-dimensional model also.
The implementation of the regularization technique is explicit in one 
dimensional case.
This makes it easy to see how our technique allows one to join 
smoothly classical solutions relevant to the tunneling and allowed
transitions.

In quantum mechanics of one degree of freedom only 
one variable $X(t)$ is present, which describes motion of a particle with mass $m=1$
through a potential barrier $U(X)$. The motion is free in the asymptotic 
regions $X\to \pm\infty$. The semiclassical 
calculation of the tunneling exponent is  performed by solving the 
classical equation of motion
$$
\frac{\delta S}{\delta X(t)} = 0
$$
 on the contour ABCD in complex time plane,
with the conditions that the solution is real in the asymptotic past (region A),
and asymptotic future (region D). The relevant solutions tend to $X\to -\infty$
and $X\to + \infty$ in regions A and D, respectively. The auxiliary parameter $T$
is related to the energy of incoming state by the requirement that the
energy of the classical solution equals to $E$. The exponent for
the transition probability is
\begin{equation}
F = 2 \Imag S - E T\;.\label{Fanswer}
\end{equation}
One notes the resemblance of these boundary conditions to the ones on the
tunneling coordinate $X$ in two-dimensional system.

In quantum mechanics of one degree of freedom, the contour ABCD may be
chosen in such a way that the points B and C are the turning points of
the solution.  Then the solution is real also at the part BC of the
contour.  Indeed, a real solution at the part BC of the contour
oscillates in the upside-down potential, $T/2$ equals to half-period
of oscillations, and the points B and C are the two different turning
points, $\dot{X} = 0$.
The continuation of this solution, according to the equation
of motion, from the point C to the positive real times corresponds to the 
real-time motion, with zero initial velocity,
towards $X\to+\infty$; the coordinate $X(t)$ stays real on the part CD of the 
contour. Likewise, the continuation back in time from the point B leads to real
solution in the part AB of the contour. In this way the reality conditions 
are satisfied at A and D. 
The only contribution to $F$
comes from the Euclidean part of the contour, and 
one can check that the expression~\eqref{Fanswer} reduces to 
\begin{equation}\label{Fanswer1}
  F(E) = 2 \int\limits_{X_B}^{X_C}\! \sqrt{2(U(X)-E)}\, dX \;,
\end{equation}
which is the standard WKB result.

The solutions appropriate for the classically forbidden and classically
allowed transitions apparently belong to different branches. 
As the energy approaches the height of
the barrier $U_0$ from below, the amplitude of the oscillations in the
upside-down potential decreases, while the period $T$ tends to a
finite value determined by the curvature of the potential at its
maximum. On the other hand, the solutions for $E > U_0$ 
always run along the real time axis, so the parameter
$T$ is always zero. Hence, the relevant solutions do not merge at $E =
U_0$, and $T(E)$ has a discontinuity at $E = U_0$. 
The regularization technique of Sec.~\ref{sec:reg_forb}
removes this discontinuity and allows for smooth transitions through the point
$E = U_0$.
The only difference with quantum mechanics of multiple degrees of
freedom is that in the latter case the bifurcation points exist  not only at the boundary of the region of classically allowed transitions, 
but also well inside the  region  of classically forbidden transitions
(but still at $E > E_S$, see Introduction and
Sec.~\ref{sec:over-barrier}).

To illustrate the situation, let us consider an exactly solvable model with 
\begin{equation*}
U(X) = \frac{1}{\cosh^2 X}\;.
\end{equation*}
We implement our regularization technique by formally changing the potential
\begin{equation}\label{Uintnew1D}
  U(X) \to \e^{- i \epsilon} U(X)
  \;,
\end{equation}
which leads to the corresponding change of the classical equations of
motion. Here $\epsilon$ is a real regularization parameter, the smallest 
parameter in the model. At the end of the calculations one takes the limit 
$\epsilon\to 0$.

We do not change the boundary conditions in our regularized classical problem,
i.e., we still require that $X(t)$ be real in the asymptotic future on
the real time axis and that $X(t')$ be real as $t'\to -\infty$ on the
part A of the contour ABCD. Then the conserved energy is real. The
sphaleron solution $X(t) = 0$ has now \emph{complex} energy (because
the potential is complex). Hence, the solutions to our classical
boundary value problem necessarily avoid the sphaleron, and one may
expect that the solutions behave smoothly in energy.

The general solution to the regularized problem is 
\begin{equation*}
  \sqrt{\frac{E}{\e^{-i \epsilon }-E}}\sinh X
  = - \cosh\left(\sqrt{2 E}(t-t_0)\right)\;,
\end{equation*}
where $t_0$ is the integration constant.  The value of $\Imag t_0$ is fixed
by the requirement that $\Imag X=0$ at positive time $t\to +\infty$,
\begin{equation*}
  \Imag t_0= \frac{T}{2} - \frac{1}{2\sqrt{2E}}
\mathrm{arg}[\mathrm{e}^{-i\epsilon} - E]\;.
\end{equation*}
The residual parameter $\Real t_0$ represents the real--time translational
invariance present in the problem.
The condition that the coordinate $X$ is real on the initial part AB
of the contour gives the relation between $T$ and $E$,
\begin{equation}\label{TEThetaE1D}
  \frac{T}{2} = \frac{1}{\sqrt{2E}}\left\{
    \pi+\arg\left(\mathrm{e}^{-i\epsilon}-E\right) \right\}\;.
\end{equation}
For $\epsilon=0$ and $E<1$, the original unregularized result 
$T/2= \pi/\sqrt{2E}$ is reproduced.

Let us analyze what happens in the regularized case in the vicinity
of the would-be special value of energy, $E = E_{\mathrm{S}}\equiv 1$. It is clear from 
Eq.~\eqref{TEThetaE1D} that T is now a 
smooth function of $E$. Away from $E = 1$,
Eq.~\eqref{TEThetaE1D} can be written as follows,
\begin{equation}\label{TEthetaE_small}
  \frac{T}{2} 
  = \left\{\begin{array}{ll}
      \displaystyle\frac{\pi}{\sqrt{2E}} & \text{forbidden region, }
1-E \gg \epsilon \\
      \displaystyle\frac{\epsilon }{\sqrt{2 E}(E-1)} & 
\text{allowed region, }  E-1 \gg \epsilon.
    \end{array}\right.\hspace{-0.7em}
\end{equation}
Deep enough in the region of forbidden transitions,
when $1-E\gg \epsilon$, the
argument in equation~(\ref{TEThetaE1D}) is nearly zero and we return
to the original tunneling solution.  When $E$ crosses the region of
size of order $\epsilon$ around $E = 1$, the argument rapidly changes
from $O(\epsilon)$ to $-\pi$, so that $T/2$ changes from
$\pi/\sqrt{2}$ to nearly zero.  Thus, at $E > 1$ we arrive to a
solution which is very close to the classical over-barrier transition,
and the contour is also very close to the real axis.  This is shown in
Fig.~\ref{fig15}. We conclude that at small but finite
$\epsilon$, the classically allowed and classically forbidden transitions
merge smoothly.

At ${E<1}$, the limit $\epsilon \to 0$ is straightforward.  For
${E>1}$ a somewhat more careful analysis of the limit ${\epsilon
\to0}$ is needed.  From Eq.~\eqref{TEthetaE_small} one observes that
the limit ${\epsilon \to0}$ with constant finite $T<\pi\sqrt{2}$ leads
to solutions with $E=1$.  Classical over-barrier solutions of the
original problem with $E > E_S\equiv 1$ are obtained in the limit
$\epsilon \to0$ provided that $T$ also tends to zero while $\tau =
T/\epsilon$ is kept finite.  Different energies correspond to
different values of $\tau$.  And that is what one expects---classical
over--barrier transitions are described by the solutions on the
contour with $T\equiv0$.


\bibliographystyle{h-physrev4}
\bibliography{slac,paper}

\end{document}